\documentclass[10pt,preprint,numberedappendix]{aastex}

\usepackage{natbib}
\usepackage{graphicx}
\usepackage{epstopdf}
\usepackage{epsfig}
\usepackage{amsmath}
\usepackage{morefloats}
\usepackage{url}

\def\gs{\mathrel{\raise0.35ex\hbox{$\scriptstyle > $}\kern-0.6em
\lower0.40ex\hbox{{$\scriptstyle \sim$}}}}
\def\ls{\mathrel{\raise0.35ex\hbox{$\scriptstyle <$}\kern-0.6em
\lower0.40ex\hbox{{$\scriptstyle \sim$}}}}

\newcommand{\um}{\,$\mu$m}

\newcommand{\lsun}{\,$\rm{L}_{\odot}$}
\newcommand{\msun}{\,$\rm{M}_{\odot}$}
\newcommand{\asec}{$^{\prime\prime}$}

\slugcomment{DRAFT \today}

\shorttitle{SEDs of 24\um-bright sources }
\shortauthors{Sajina et al.}

\begin{document}

\title{{\sl Spitzer} and {\sl Herschel}-based SEDs of 24\um-bright $z$\,$\sim$\,0.3\,--\,3.0 starbursts and obscured quasars}
\author{Anna Sajina$^1$, Lin Yan$^2$, Dario Fadda$^3$, Kalliopi Dasyra$^{4}$, Minh Huynh$^5$}

\affil{$^1$ Department of Physics \& Astronomy, Tufts University, Medford, 02155, MA, USA}
\affil{$^2$ Infrared Processing and Analysis Center, California Institute of Technology, Pasadena, CA 91125, USA}
\affil{$^3$ NASA {\sl Herschel} Science Center, California Institute of Technology, Pasadena, CA, 91125, USA}
\affil{$^4$ Observatoire de Paris, LERMA (CNRS:UMR8112), 61 Av. de l\'\,Observatoire, F-75014, Paris, France}
\affil{$^5$ International Center for Radio Astronomy Research, M468, University of Western Australia, Crawley, WA 6009, Australia}

\begin{abstract}

In this paper, we characterize the infrared spectral energy distributions (SEDs) of mid-IR selected $z$\,$\sim$\,0.3\,--\,3.0 and $L_{IR} \sim 10^{11} - 10^{13}L_\odot$ galaxies, and study how their SEDs differ from those of local and high-$z$ analogs. Infrared SEDs depend both on the power source (AGN or star-formation) and the dust distribution. Therefore, differences in the SEDs of high-$z$ and local galaxies provide clues as to differences in their physical conditions. Our mid-IR flux-limited sample of 191 sources is unique in size, and spectral coverage, including {\sl Spitzer} mid-IR spectroscopy. Here we add {\sl Herschel} photometry at 250\um, 350\um, and 500\um, which allows us, through fitting an empirical SED model, to obtain accurate total IR luminosities, as well as constrain the relative contributions of AGN and starbursts to those luminosities. Our sample includes three broad categories of SEDs: $\sim$\,23\% of the sources are AGN (i.e. where the AGN contributes $>$\,50\,\%\ of $L_{\rm{IR}}$), $\sim$\,30\,\%\ are starbursts where AGN contributes $<$\,20\,\%\ of $L_{\rm{IR}}$ and the mid-IR spectra are starburst-like (i.e. strong PAH features); and the largest group ($\sim$\,47\%) are composites which show both significant AGN and starburst activity. The AGN-dominated sources divide into ones that show a strong silicate 9.7\um\ absorption feature, implying highly obscured systems, and ones that do not. The high-$\tau_{9.7}$ sources are half of our  $z$\,$>$\,1.2 AGN, but show SEDs that are extremely rare among local AGN. The 30\%\ of the sample that are starbursts, even the $z$\,$\sim$\,2, $L_{\rm{IR}}$\,$\sim$\,$10^{13}$\lsun\ ones, have lower far-IR to mid-IR continuum ratios than local ULIRGs or the $z$\,$\sim$\,2 sub-mm galaxies -- effectively the SEDs of our $z$\,$\sim$\,2 starburst-dominated ULIRGs are much closer to those of local LIRGs than ULIRGs. This is consistent with our earlier finding that, unlike local ULIRGs, our high-$z$ starbursts are typically only in the early stages of a merger. The SEDs of the composite sources are most similar to the local archetypal warm ULIRG, Mrk231, which supports the interpretation of their consisting of both AGN and starburst components. In summary, our results show that there is strong evolution in the SEDs between local and $z$\,$\sim$\,2 IR-luminous galaxies, as well as that there is a wide range of SEDs among high redshift IR-luminous sources. The publicly-available SED templates we derive from our sample will be particularly useful for infrared population synthesis models, as well as in the interpretation of other mid-IR high-$z$ galaxies in particular those detected by the recent all sky {\sl WISE} survey. 
\end{abstract}


\section{Introduction \label{sec:intro}}

Understanding the nature of dusty galaxies at redshifts  $z$\,$\sim$\,1\,--\,3 is key to the study of galaxy evolution, since this is when the star-formation rate (SFR) density of the universe peaked \citep[e.g.][]{bouwens11}, when most of the stars we see in the local Universe were formed \citep[e.g.][]{marchesini09}, as well the epoch of peak quasar number density  \citep{wall05,richards06}. Along with the M-$\sigma$ relation \citep{kormendy95,magorrian98,tremaine02,shankar12}, this common peak activity epoch suggests that the growth of galaxies is intimately linked with the growth of their central supermassive black holes. IR-luminous galaxies, in particular Luminous Infrared Galaxies (LIRGs, defined as having $L_{\rm{IR}}$ in the range $10^{11}$\,--\,$10^{12}$\lsun) and Ultra Luminous Infrared Galaxies \citep[ULIRGs, defined as having $L_{\rm{IR}}$\,$>$\,$10^{12}$\lsun, for a review see][]{sm96review,lonsdale06review}, are particularly important since they increase dramatically in number density from today until $z$\,$\sim$\,2, leading to a strong IR luminosity function evolution, which makes them the dominant contributor to the SFR density peak  \citep{lefloch05,caputi07}. In addition, theory suggests that ULIRGs and quasars are directly linked, with the late stages of a major mergers leading to the high SFR, high dust obscuration ULIRG phase, followed by a quasar phase \citep{sanders88,hopkins08}.  This scenario is well supported in the local universe \citep[e.g.][]{surace00,veilleux02,canalizo07}. However, there are indications that the ULIRGs at $z$\,$\sim$\,2 are not analogous to those found locally. In particular, the high-$z$ ULIRGs show colder characteristic dust temperatures \citep{chapman04_smgs,sajina06,pope06,huynh07_smg,muzzin10,seymour10,mrr10}; higher molecular gas fractions \citep{tacconi10,yan10}; and, unlike local ULIRGs, are often found in only the early stages of a merger or even in isolated disks \citep[e.g.][]{forster-schreiber09,narayanan10,engel10,zamojski11}.

An important tool in addressing the evolution of the IR-luminous population is the infrared spectral energy distribution (SED) which depends on both the relative strength of the AGN and the star-formation activity, as well as dust distribution. As an example, SEDs that peak at longer wavelengths (i.e. cooler dust temperatures) are believed to be indicative of either isolated galaxies or galaxies in the early stages of a merger, while warmer dust temperatures are indicative of galaxies in the late stages of a merger \citep[e.g.][]{hayward12}.  Indeed, a key finding that high redshift ULIRGs are indeed not like local ones is that they tend to show colder dust temperatures (see above), although this finding is based exclusively on far-IR/sub-mm selected samples. Galaxies with stronger mid-IR continua are indicative of stronger AGN activity, while galaxies with strong polycyclic aromatic hydrocarbon (PAH) features in their mid-IR spectra are indicative of largely starburst galaxies \citep[e.g.][]{genzel98,laurent00,tran01,veilleux09}. Therefore, characterizing the SEDs of high-$z$ ULIRG populations tells us of their power source and overall dust geometry, while characterizing how these high-$z$ ULIRG SEDs differ from those of local ULIRGs tells us how such fundamental properties evolve with redshift. Infrared SEDs  are also an essential ingredient in galaxy evolution models \citep[e.g.][]{lagache03,valiante09,leborgne09,bethermin11}. Current models, however, have two key limitations in their SED treatment: they assume that SED templates derived locally are directly applicable at high redshift (i.e. no SED evolution), and they either completely neglect the role of AGN or adopt a single AGN template \citep{franceschini01,valiante09}. These limitations arise because, until recently sufficiently good spectral coverage, for large, well defined samples of high-$z$ sources has not been available hence deriving SED templates appropriate for $z$\,$\sim$\,2 starburst or AGN sources have not been possible. Starting from mid-IR selected samples helps because this selection results in samples that include both AGN and starbursts. Characterizing the overall SEDs of mid-IR selected sources is important for galaxy evolution studies since half of the Cosmic Infrared Background at its peak ($\sim$\,70\,--\,160\um) is resolved by sources with $F_{24}$\,$>$\,0.2\,mJy \citep{dole06}.  Mid-IR-based SED templates are important in the interpretation of the high redshift mid-IR bright sources detected by the recent all sky {\sl WISE} \citep[{\sl Wide-field Infrared Survey Explorer};][]{wright10} survey, especially at 22\um. Beyond the generation of templates, understanding the nature of the mid-IR selected sources (specifically the role of AGN therein), requires the availability of mid-IR spectra since this regime is largely dominated by the PAH and silicate absorption features, which cannot be distinguished with broadband data alone. 

Our group has been involved in a detailed multi-wavelength study of an exceptional sample of 191 24\um-selected sources with mid-IR spectra as well as extensive multi wavelength coverage from the X-ray to the radio including {\sl HST} NICMOS imaging \citep{yan07,sajina07a,sajina07b,sajina08,dasyra09,sajina09,bauer10,yan10,zamojski11}.  Some key conclusions include: 1) the bulk of this sample appear AGN-dominated using mid-IR spectral diagnostics, although $\sim$\,30\% are starburst-dominated including some $\sim$\,$10^{13}$\lsun, $z$\,$\sim$\,2 sources; 2) where X-ray data are available, our mid-IR AGN are not individually detected, suggesting potentially Compton-thick AGN; 3) the bulk of our sample shows signs of mergers/tidal interactions; and 4) like the SMGs, the small number of our sources with CO measurements suggest a higher molecular gas fraction than seen in local ULIRGs \citep{yan10}. Ultimately however, our previous studies on the nature of these sources have been limited by our incomplete knowledge of their overall infrared luminosities. 

In this paper, we constrain the full IR SEDs of these mid-IR selected sources in order to determine accurate total IR luminosities for our sources, and the fractions of $L_{\rm{IR}}$ that are due to AGN/star-formation activity. We address how well do mid-IR based AGN/starburst classifications translate to the overall IR SED. Using our sample which is exceptional in size, and spectral coverage, we produce SED templates appropriate for high redshift starburst and obscured quasar systems. We address how the SEDs of galaxies of a given luminosity evolve with redshift by comparing our SED templates with other IR SED templates based on sources of comparable luminosity and/or redshift. Our study is made possible in particular thanks to the observations of the First Look Survey Field with {\sl Herschel}'s Spectral and Photometric Imaging REceiver \citep[SPIRE;][]{griffin10} operating at 250\um, 350\um, and 500\um, as part of the {\sl Herschel} Multi-tiered Extragalactic Survey \citep[HerMES;][]{oliver10}. Throughout this paper we adopt the 7-year WMAP cosmological parameters, specifically $\Omega_{M}$\,=\,0.274, $\Omega_{\Lambda}$\,=\,0.725, and $H_0$\,=\,70.2\,km\,s$^{-1}$\,Mpc$^{-1}$ \citep{komatsu11}.

\section{Data}

\subsection{Sample selection\label{sec_sample}}

A number of programs, the largest of which is by our own group, involve {\sl Spitzer} IRS spectra of 24\um-bright sources in the {\sl Spitzer} Extragalactic First Look Survey\footnote{http://ssc.spitzer.caltech.edu/fls/}  (xFLS) field.  We combine our IRS data with archival data to construct an xFLS "IRS supersample" of 191 sources. The criteria for this supersample are: 1) to be located in the inner 2.7\,sq.deg. of the xFLS; 2) to have a 24\um\ flux of $F_{24}$\,$>$\,0.9\,mJy; and 3) to have an R magnitude of $R$\,$\geq$\,20. The bulk of the IRS sample comes from our {\sl Spitzer} GO2 program \citep[see][for details]{dasyra09}, followed by our {\sl Spitzer} GO1 sources \citep{yan07,sajina07a}. An additional 17 sources from several different programs\footnote{Our supersample includes four IRS sources from PID\#20128 (PI Lagache), two sources from PID\#30447 (PI Fazio), and one source from PID\#20542 (PI Borys) that to our knowledge have not been published to date. } \citep{weedman06,lacy07_letter,ams08} also meet our selection criteria. Combined, these samples constitute a `supersample' of 212, of which, 191 have redshifts (see Section\,\ref{sec_redshifts}). In this paper, we only consider the sources with redshifts.

\begin{figure}[!h]
\includegraphics[scale=0.37]{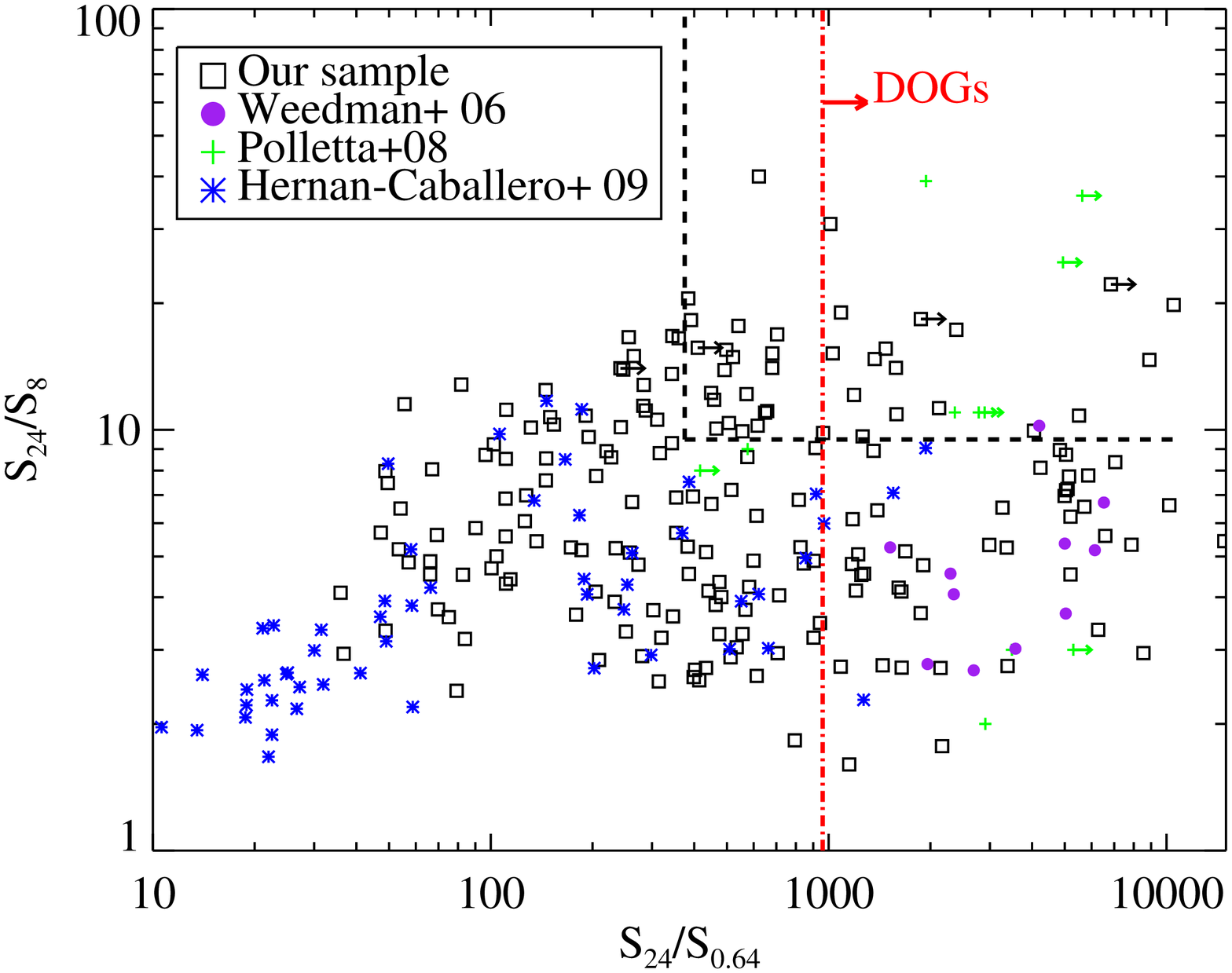}
\includegraphics[scale=0.37]{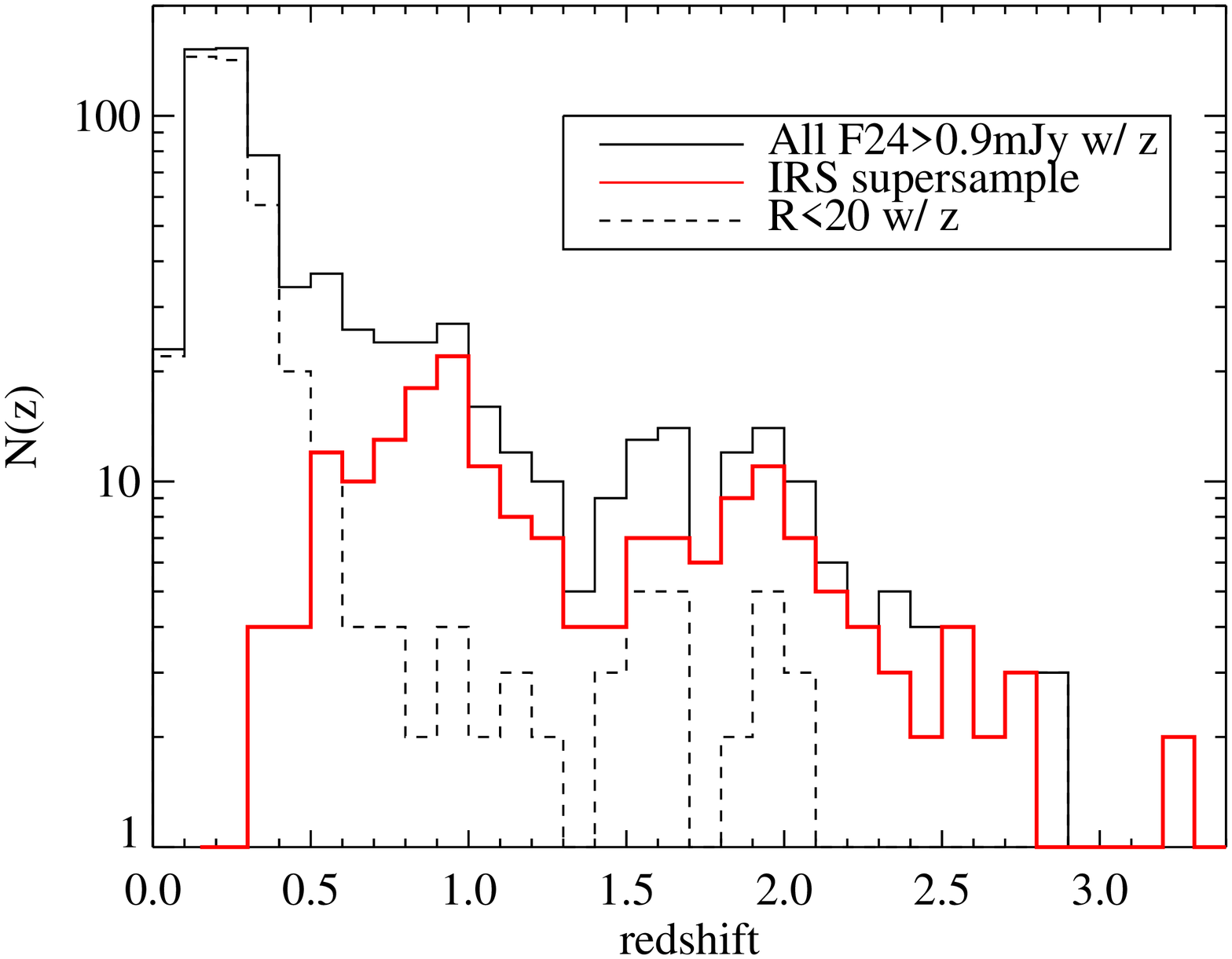}
\caption{{\it Top:} Our sample in the optical-infrared selection plot compared with similar samples from the literature. The dashed lines indicate our GO1 sample's color selection. The red dot-dashed line indicates the "dust obscured galaxies" or "DOGs" selection \citep{dey08}. {\it Bottom:} The redshift distribution of our total IRS sample of 191 sources. We also plot the redshift distribution of the $R<20$, $F_{24}$\,$>$\,0.9\,mJy sources which are excluded from our sample.  \label{fig_selection}}
\end{figure}

Our  IRS supersample contains just under half the xFLS sources that meet the above photometric criteria; however it is representative of this parent sample for $z$\,$>$\,1 and $R$\,$>$\,20 sources.  Specifically, the IRS sample has the same $F_{24}$/$F_{8}$ color distribution as the parent sample, is essentially complete for the $R$\,$>$\,22 sources, but is incomplete in the $R$\,=\,20\,--\,22 optical magnitude range. Fig\,\ref{fig_selection}{\it top} shows the color distribution of our sample in $F_{24}$/$F_{0.64}$ and $F_{24}$/$F_{8}$ compared with related samples from the literature. Fig\,\ref{fig_selection}{\it bottom} shows the redshift distribution of our sample (see Section\,\ref{sec_redshifts} for details) compared with the redshift distribution of all xFLS $F_{24}$\,$>$\,0.9\,mJy sources with available redshifts, based on the redshift surveys of \citet{papovich06} and \citet{marleau07} as well as SDSS redshifts. To illustrate the effect of our $R>20$ selection, we separately show the redshift distribution of the $R<20$ sources, $\sim$\,70\% of which have known redshifts.  The primary effect of our optical brightness cut is to exclude the $z$\,$\sim$\,0.2 peak which is likely dominated by normal spiral galaxies. The second effect is to exclude Type 1 AGN at all redshifts. Based on their optical spectral classification \citep{papovich06},  the optically-bright sources at $z$\,$>$\,1 are all broad line QSOs.  Our incompleteness in the $R$\,=\,20\,--\,22 range is likely to affect predominantly the $z$\,$\sim$\,0.5\,--\,1.0 range. At higher redshifts, $z$\,$\sim$\,1\,--\,3, our sample is representative of a pure 24\um\ flux limited survey. 

\subsection{Redshifts \label{sec_redshifts}}
The bulk of the redshifts used here come from the IRS spectra themselves \citep[see][]{yan07,sajina07a,dasyra09}.  The high-confidence redshifts are typically those based on clear PAH features and have uncertainties ($\delta z$) in the range 0.01-0.03 \citep{dasyra09}.  Redshifts based only on the silicate absorption feature have typical uncertainties of $\sim$\,0.1 up to 0.2 \citep{sajina07a}.  Most of our lower-$z$ sources have redshifts from \citet{papovich06} or \citet{marleau07}.  We also have optical spectroscopic redshifts based on targeted Keck and Gemini spectral follow-up \citep[see e.g.][]{choi06,yan07,sajina08}. Several of the sources are among the spectroscopic sample of mid-IR-selected AGN sources presented in \citet{lacy07}. Altogether, 69 of our sources have optical spectroscopic redshifts, which are found to be in good agreement with the IRS spectroscopic redshifts. We adopt the optical spectroscopic redshift whenever available. Nine of the sources do not have optical spectroscopic redshifts, and have IRS spectroscopic redshifts based on weak and uncertain features, and two sources have discrepant mid-IR and optical redshifts. All redshifts are listed in Table~\ref{table_prelims} where the 11 uncertain redshifts have a question mark beside them. 

\subsection{{\sl Spitzer} IRS diagnostics: PAH equivalent widths and $\tau_{9.7}$ \label{sec_mir}}

All mid-IR IRS spectra were fit with the approach adopted in \citet{sajina07a}. This is a simple empirical model involving a power law 5-15\um\ continuum, a Galactic Center mid-IR extinction curve \citep{chiar06}, and a PAH template derived from the local starburst galaxy NGC7714. This allows us to determine a continuum level and the silicate feature depth, $\tau_{9.7}$. The PAH equivalent widths are determined from fitting Lorentz profiles of the individual 3.3, 6.2, 7.7, 8.6, and 11.3\um\ PAH features onto the previously determined continuum.  Our approach was originally devised as a way to deal with noisier data over a range of redshifts and hence varying rest-frame coverage. Some caveats include: 1) across our redshift range the 7.7\um\ PAH feature is covered by the largest fraction of the sources (hence is the one we usually use), but for the $z$\,$<$\,0.9 sources without IRS SL data \citep[see][]{dasyra09}  our 7.7\um\ equivalent widths come from the PAH template fit in the first step of the process (see above); 2) the continuum beyond the silicate absorption feature is poorly constrained for $z$\,$>$\,2.2 sources, giving large uncertainties on $\tau_{9.7}$, and 3) for sources with strong PAH there is a degeneracy between the PAH features strength and the 9.7\um\ silicate feature depth. Our fitting method tends to give larger silicate optical depths for strong PAH sources than other approaches \citep[see][for a direct comparison]{sajina09}.  This approach estimates the depth of the silicate feature relative to an unextincted continuum which is always larger than an estimate of the depth of the feature relative to the observed continuum. For our adopted extinction curve, the latter can be obtained by dividing our $\tau_{9.7}$ values by 1.4 \citep[see][for further discussion]{sajina09}. Table\,\ref{table_prelims} gives the best-fit 7.7\um\ feature equivalent widths of the 7.7\um\ feature as well as the $\tau_{9.7}$ values for our sources. Throughout, we follow the convention of \citet{sajina07a}, and use EW7.7\,$>$\,0.9\um\ as the definition of a "strong PAH" source, which are sources dominated by star-formation in the mid-IR. We also define "high-$\tau_{9.7}$" sources as those sources that have $\tau_{9.7}$\,$>$\,1. 

\begin{deluxetable}{ccc}
\tablecolumns{3}
\tablewidth{3in}
\tabletypesize{\scriptsize}
\tablecaption{\label{table_det} SED coverage statistics}
\tablehead{\colhead{$\lambda_{\rm{obs}}$[\um]\tablenotemark{a}} & \colhead{\% detected (\#det/\#obs)} & \colhead{Est. confusion\tablenotemark{b}}}
\startdata
0.64 & 89 (170/191)  & \\
3.6 & 94 (180/191) & $\sim$\,13\% \\
4.5 & (/191) & $\sim$\,13\% \\
5.8 & (/191) & $\sim$\,13\% \\
8.0 & 96  (184/191) & $\sim$\,13\% \\
24 & 100 & \\
71.4 &  69 (132/191) & $\sim$\,12\% \\ 
155.9 & 18(35/191) & $\sim$\,28\% \\
250 & 60(114/191) & $\sim$\,13\% \\
363.0 & 38(79/191) & $\sim$\,19\% \\
517.0 & 12(31/191)& $\sim$\,28\% \\
1200\tablenotemark{c} &  20 (10/51) &  $\sim$\,2\% \\
1.4GHz & 59 (113/191)  & \\
610MHz & 37 (71/191) &  
\enddata
\tablenotetext{a}{We list the actual central wavelengths, although in the text we refer to the more common band names: e.g. MIPS160 instead of 155.9. The SED fits use the instrumental filters.}
\tablenotetext{b}{These are approximate estimates for the fraction of sources that may suffer from confusion (multiple sources contributing to the flux/uncertain ID) (see text for details).}
\tablenotetext{c}{This includes one source which is detected with SCUBA at 850\um. As a constraint on the SED, this is close enough to the MAMBO observations to be included here. }
\end{deluxetable}

\subsection{Optical/near-IR photometry}
The $R$ band data comes from the KPNO Mosaic-1 image of the xFLS \citep{fadda04}. The 5\,$\sigma$ limit of this survey is $R$\,$=$\,25.5 (Vega).  By selection, all our sources have $R$\,$\geq$\,20. A total of 23 sources (12\%) are undetected in $R$, and hence we adopt the above 5\,$\sigma$ limit. 

The IRAC 3.6, 4.5, 5.8, and 8.0\um\ fluxes for our sample come from the IRAC map of the xFLS, where the 6\arcsec\ aperture rms values are 2.3, 3.2, 15, and 14.4$\mu$Jy for the four bands respectively \citep{lacy05}.  For about 10\% of the cases, the IRAC counterpart to the MIPS source is ambiguous (see Appendix\,\ref{sec_irac_blend} for details). Here we adopt the IRAC id's given in \citet{dasyra09} and \citep{sajina07a}.  The $R$-band and IRAC flux densities of our sources are given in Table\,\ref{table_fluxes}.

\subsection{{\sl Spitzer} MIPS data}
Our 24\um\ flux densities are drawn from the \citet{fadda06} catalog based on the MIPS 24\um\ image of the xFLS field. The flux errors are typically $\sim$\,0.04-0.16\,mJy. The xFLS also has a MIPS70\um\ and MIPS160\um\ scanmap images presented in \citet{frayer06} along with the associated 7\,$\sigma$ point source catalogs. For the 70\um\ image the typical 1\,$\sigma$ noise is 2.8\,mJy in the main field, and 1.6\,mJy in the verification field. For the 160\um\ image the noise varies significantly across the field, but has typical 1\,$\sigma$ values of $\sim$\,10\,mJy in the smaller verification field and $\sim$\,20\,mJy in the main field. For our GO1 sample sources not in the \citet{frayer06} 70\um\ catalog, we obtained MIPS\,70\um\ targeted photometry that reaches a depth comparable to the verification field (details on the observing strategy and data reduction for the GO1 sample are given in \citet{sajina08}). In {\sl Spitzer} GO4, we also observed the GO2 sources without detections in the xFLS scanmap.  All existing MIPS 70\um\  (scan-map and targeted photometry) were co-added in quadrature. The fluxes and their uncertainties were estimated from PRF photometry on this co-added image \citep[using APEX;][]{makovoz05}. \citet{frayer09} point out that a multiplicative factor of 1.2 needs to be applied to their earlier xFLS data. Since this factor is largely due to the PRF model used (same as we use here), we apply this factor to both to our scanmap and photometry data. To obtain 160\um\ flux densities for our sources, we ran APEX on the xFLS scanmap image keeping all 2\,$\sigma$ sources and cross-matching this list with our source positions. We apply the multiplicative correction factor of 0.97 given in  \citet{frayer09}. Ten sources have detections using targeted 160\um\ photometry \citep{sajina08}. However, we find that filtering of these small 160\um\ photometry fields leads to significant (30-50\%) flux loss. Since six of these sources are found to still have 2-3\,$\sigma$ detections in the xFLS scanmap, for the purposes of this paper, it was judged simpler to only use the scanmap data. Following \citet{stansberry07}, color-correction factors (divisive) of 0.919 for MIPS70\um, and 0.969 for MIPS160\um\ are applied (these are reliable to within $\sim$\,2\% assuming dust emission in the range 30\,--\,100\,K). The overall calibration uncertainty is 2\% for MIPS24, 5\% for MIPS70, and 12\% for MIPS160 \citep{stansberry07}.  These are added in quadrature with the local rms values to obtain the total errors. The MIPS flux densities and their associated errors are given in Table\,\ref{table_fluxes}. 

\subsection{{\sl Herschel} SPIRE data}
The xFLS field was observed with the {\sl Herschel} SPIRE instrument as part of the {\sl Herschel} Multi-tiered Extragalactic Survey (HerMES) survey.  The xFLS observations took 17.10hr for the entire field. The {\sl Herschel} SPIRE confusion limit is measured to be 5.8, 6.3 and 6.8 mJy/beam at 250, 350 and 500 microns, respectively \citep{nguyen10}. The typical 1\,$\sigma$ rms values in the xFLS in all three bands are comparable to this suggesting that they are essentially confusion limited. 

We use the publicly released level 2 maps of the xFLS field, and extract our sources' SPIRE flux densities using PSF fitting at the positions of the MIPS24\um\ sources. These SPIRE maps are calibrated with the assumption of a flat spectrum, while we assume that we are typically in the Rayleigh-Jeans part of the spectrum with $S_{\nu}\propto\nu^{3}$ corresponding to multiplicative color-corrections of 0.9070,0.9180, and 0.8952 for the three SPIRE bands respectively. Wherever we have both 250\um\ and 350\um\ detections we compute the spectral index between them and use the appropriate color correction for that spectral index. We also apply divisive pixelization corrections to the three SPIRE bands which are respectively 0.951,0.931 and 0.902. Following the SPIRE User's manual, we assume a calibration error of 7\% as well as a pixelization error of 2\% for each SPIRE band. We compute the total error as the quadrature sum of the rms, the calibration error, and the pixelization error. 

At the 3\,$\sigma$ level, 114 sources are detected in the SPIRE 250\um\ image, 79 sources are detected in the 350\um\ image, and 31 sources are detected in the 500\um\ image. However, the large beam sizes in the far-IR regime lead to significant confusion as to what degree the observed emission is due to our particular 24\um\ source. The fraction of our sources with one or more additional 24\um\ source within SPIRE FWHM/2 of a given MIPS\,24\um\ position is 13\%, 19\%, and 28\% respectively for the 250\um, 350\um, and 500\um\ beams. In Appendix\,\ref{sec_fir_blend}, we discuss the treatment of such cases.  

 \subsection{SCUBA\,850 and MAMBO\,1.2mm data}
MAMBO 1.2\,mm observations for the entire GO1 sample were presented in \citet{sajina08}. A number of additional 24\um-bright sources were presented in \citet{lutz05}. In addition, \citet{ams09} provide us with the millimeter fluxes for MIPS8392, MIPS22722, 12509696, 19456000, and 19454720. In total 50 sources have 1.2\,mm photometry, although the majority of these are non-detections. One source, MIPS8543 has a SCUBA 850\um\ flux from \citet{frayer04}. For simplicity, as this is a single source, we include its flux in the $F_{1200}$ column in Table~\ref{table_fluxes}, although of course we use its correct observed wavelength in the analysis. 

\subsection{Radio 1.4\,GHz and 610\,MHz data}
The radio data come primarily from two images of the xFLS field at 610\,MHz with the Giant Meterwave Radio Telescope (GMRT) \citep{garn07} and at 1.4GHz with the Very Large Array (VLA) \citep{condon03}. The bulk of our 1.4GHz flux densities are based on the 4\,$\sigma$ ($\simeq$\,90\,$\mu$Jy) catalog extracted from this image (Jim Condon, private comm.). A smaller, $\sim$\,1\,sq.deg. field in the center of the xFLS was imaged down to $\sigma$\,$\sim$\,8.5\,$\mu$Jy with the Westerbork Synthesis Radio Telescope (WSRT) \citep{morganti04}.  A total of 47 of our sources are detected in this deeper field. We compare the VLA and WSRT fluxes for these sources, and find a median difference (VLA-WSRT) of 0.039\,mJy with a standard deviation of 0.098\,mJy, with no strong outliers (suggesting variability is most likely not a significant issue for this sample).  This small offset could be attributed to the slightly different bandpasses. We adopt the WSRT fluxes wherever available, due to their much higher signal-to-noise, and hence reliability.  Overall, 113 sources are detected at 1.4\,GHz and 71 at 610\,MHz. The radio fluxes for our sample are given in Table\,\ref{table_fluxes}. 

\section{Analysis \label{sec_analysis}}

\subsection{Composite SED model\label{sec_model}}

We use empirical SED model fitting to determine the far-IR properties of our mid-IR selected sources. This allow us to: 1) determine the total IR luminosities, 2) estimate the relative contribution of the mid-IR (likely AGN-dominated) continuum to this total, 3) compute rest-frame colors, and 4) construct average templates.  The disadvantage of this approach is that the physical interpretation is not intrinsic to the model, but rather relies on a priori assumptions such as "the hot dust continuum originates in an AGN torus, while the cold dust continuum originates in star-forming regions" . Unfortunately, to date none of the more physically-inspired SED models are able to self-consistently handle the full range of SED types from essentially pure AGN to pure starbursts that characterizes this sample, while our empirical approach is able to characterize all  SED types found in our sample.

Our composite empirical model is given in Equation\,\ref{eq_model}, where we abreviate $\nu f_{\nu}$ with $F$:

\begin{eqnarray}
\label{eq_model}
  F&=&a_{\rm{stars}}F_{\rm{stars}}+a_{\rm{PAH}}F_{\rm{PAH}} +a_{\rm{hot}}F_{\rm{hot}}e^{-\tau_{hot,\nu}}+ \nonumber\\
 & +&a_{\rm{warm}}F_{\rm{warm}} +a_{\rm{cold}}F_{\rm{cold}}
\end{eqnarray}

The $F_{\rm{stars}}$ component, serves to account for the 1.6\um\ stellar bump (where observed) and  is based on a 2\,Gyr-old solar metallicity SSP from \citet{maraston05}.  This age was chosen to not exceed the age of the Universe for the highest redshift objects in our sample, although beyond $\sim$\,1\,Gyr, all SSPs look fairly similar in the near-IR regime which is relevant here. The PAH component is a fixed template based on the NGC7714 starburst \citep[see][]{sajina07a}. For the screen extinction on the hot component, we use the Galactic Center extinction curve of \citet{chiar06}.

\begin{figure}[ht!]
\plotone{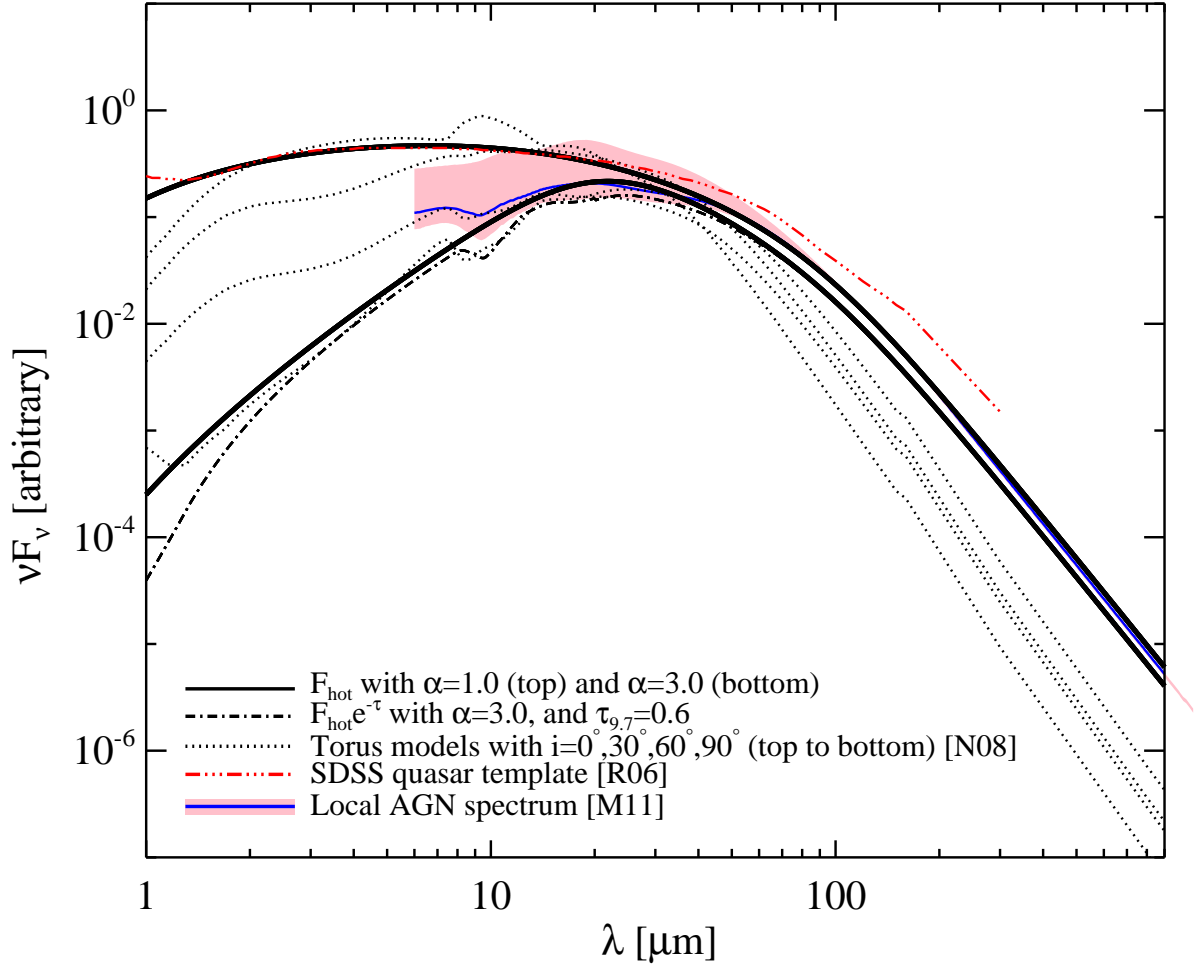}
\caption{Our $F_{\rm{hot}}$ (i.e. AGN torus) component is constrained to $\alpha$\,=\,1\,--\,3, as shown by the thick solid curves, with the effect of an dust screen ($F_{\rm{hot}}e^{-\tau}$) shown as the thick dot-dash curve. This empirical model is chosen to match the range of observed AGN SEDs (as shown here through the SDSS quasar template of \citet[][here R06]{richards06} as well as the local AGN average spectrum of \citet[][here M11]{mullaney11}). Our model is also comparable to the range of AGN torus SEDs from the radiative transfer models of \citet[][here N08]{nenkova08}. For comparison, we show the torus models for a few different inclination angles as well as overall level of obscuration (see text for details). \label{fig_agnmodels}}  
\end{figure}

The hot component in most cases is largely associated with AGN tori, although some extreme (very young) starbursts can have significant hot dust emission as well \citep{lu03,roussel03}. We adopt a broken tapered power law given by: 
\begin{equation}
\label{eq_hot}
F_{\rm{hot}}=\frac{\nu}{\big(\frac{\nu}{\nu_o}\big)^{\alpha}e^{0.5\nu}+\big(\frac{\nu}{\nu_o}\big)^{-0.5}+\big(\frac{\nu}{0.3\nu_o}\big)^{-3.0}},
\end{equation}

where $\alpha$ is the mid-IR spectral index and $\nu_o$ is the characteristic frequency, which roughly determines the location of the spectral peak.  The exponential tapering mimics the effect of sublimation on the spectrum, while $\nu^{-3}$ is effectively the RJ tail of a $\beta$\,=\,1 dust component, and the flatter component in the denominator is merely a means of softening this double power law peak \citep{mullaney11}. Figure\,\ref{fig_agnmodels} shows a comparison of our empirical hot component with a selection of clumpy torus models \citep{nenkova08}.  Here, we look at the models that result from combinations of parameters, that are most consistent with observations \citep[see][]{nenkova08}. Specifically, we fix the maximal radial extent at 30 times the sublimation radius, the opening angle at 30$^{\circ}$, the slope of the radial density distribution at 2. The torus models shown in Figure\,\ref{fig_agnmodels} vary in two important parameters: 1) the inclination angle; and 2) the overall optical depth. We find that the mid-IR slope can be described to vary from $\alpha$\,$\sim$\,1 for lower opacity, face-on tori to $\alpha$\,$\sim$\,3 for higher opacity, edge-on models (see Figure\,\ref{fig_agnmodels}). The average, intrinsic AGN SED in \citet{mullaney11}, is consistent with this range. Type-1 quasar templates \citep[e.g.][]{richards06} are close to the $\alpha$\,=\,1, no screen extinction, hot component. In our fitting, the slope, $\alpha$, is restricted to the range 1-3, while $\nu_o$ varies such that the hot component peak is in the range $\sim$\,20\,--\,40\um, effectively a function of the radial extent of the torus.  The silicate absorption feature is introduced with a screen extinction as seen in Figure\,\ref{fig_agnmodels}.

The far-IR emission, is described by the cold component, for which we use the form $F_{\nu}$\,$\propto$\,$(1-\exp(-(\nu/\nu_o)^{\beta})B(\nu,T)$. This is a generalized form that reduces to the more commonly used $\propto$\,$B(\nu,T)\nu^{\beta}$ in the optically-thin regime \citep[see][for a discussion]{hayward11}.  The free parameters are an overall amplitude, $\nu_o$, $\beta$, and $T$; however, some of those are held fixed in cases of poor far-IR coverage (see Section\,\ref{sec_fits}).  We allow temperatures between 10 and 100\,K, spectral indices, $\beta$ between 1.0 and 2.5, and $\nu_o$ such that the transition to the optically-thin regime is in the range 50-300\um. This approach aims to describe the far-IR peak of the SED; however, we do not go further at interpreting the derived parameters as there are strong degeneracies between them\footnote{For a description of the $T$-$\beta$ degeneracy see \citet{sajina06}.}. 

The warm component is given by $F_{\rm{warm}}$\,=\,$\nu^{1-\alpha_{\rm{w}}}e^{-\nu_o/\nu}$. The power law represents the emission of stochastically-heated very small grains (VSG) \citep{desert90}. We do not have sufficient data to fit more than an overall normalization and hence we fix  $\alpha_{\rm{w}}$\,=\,4, and set $\nu_o$ such that the peak of the warm component is at $\sim$\,50\um. We try various options here and find that these choices work well for our sample.  They are still somewhat arbitrary, largely motivated by the Galactic cirrus VSG component in \citet{desert90}, but serve the purpose of providing a smooth transition between the hot and cold components without too much competition with either.

\subsection{SED fits \label{sec_fits}}
We fit the above composite model to the rest-frame 1\,--\,1000\um\ of all our sources using a Markov Chain Monte Carlo (MCMC) code \citep[see][ and Appendix\,\ref{sec_mcmc}]{sajina06}, and adopting the lowest $\chi^2$ solution as our best-fit. Along with the broadband photometry, the fit includes the IRS spectra, convolved with a series of artificial filters that give observed frame fluxes at 16,18, 20, 22, 26, and 28\um\ (24\um\ is already given by the MIPS24\um\ flux). These `filters' are all square with $\Delta\lambda$\,=2\um\ \citep[see e.g.][]{hc09}.  Only formal detections are used in the fits (typically 3\,$\sigma$, but we allow 2\,$\sigma$ SPIRE\,250\um\ and 350\um\ photometry). The mid-IR part of the spectra are well sampled due to the IRS spectra; however, some sources have non-detections in the near-IR (IRAC) bands, or the far-IR (160-500\um) bands. For sources of known weak PAH emission ($EW7.7$\,$<$\,0.9\um), we do not fit a PAH component. For sources with less than two IRAC detections, we do not fit a stellar component. For sources without far-IR detections, the cold component is fixed to an optically-thin template of fixed temperature and $\beta$ (usually $T$\,=\,50\,K, $\beta$\,=\,1.5) where only the amplitude, $a_{\rm{cold}}$ is left as a free parameter. Therefore, the number of free parameters ranges from 5 for sources with poor coverage in the near-IR and far-IR to 9 for sources with maximum coverage.  Lastly, only for the sources without far-IR detections, we impose a $\chi^2$ penalty to solutions that exceed the upper limits in the far-IR\footnote{A similar restriction is not required for sources without IRAC detections, since in this case, we do not fit a stellar component at all, and the SED models are essentially always below the IRAC upper limits.}. Without such a penalty, best-fit solutions with unjustifiably high IR luminosities can be found.
  
\begin{figure*}[t!]
\includegraphics[scale=0.35]{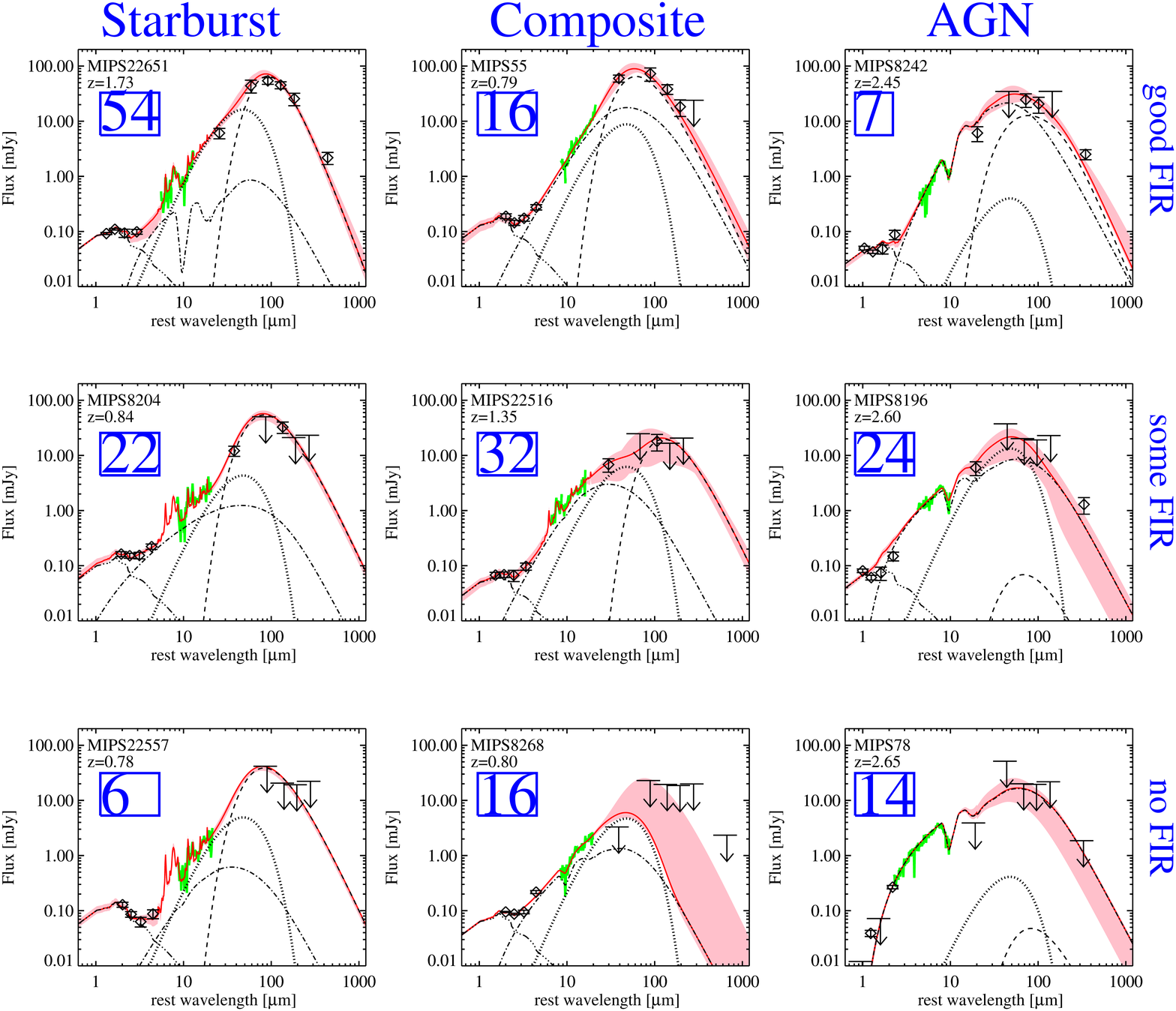}
\caption{Examples of individual SED fits representative of the range of SED types (left-to-right) and far-IR coverage (top-to-bottom), where far-IR means observed 70\um\ and long-ward. The boxed numbers represent the number of sources in each category. The solid red curves show the best fits and the pink shaded area represents the spread of solutions within $\chi^2_{min}+1$. The dashed-triple dot curves represent the stellar component, the dot-dash curves are the hot component, the dotted curves are the warm component, and the dashed curves represent the cold component. \label{fig_seds}}  
\end{figure*}

In Figure\,\ref{fig_seds}, we show examples of what the SED fits look like for different types of source and different far-IR coverage. Here the left-hand column shows strong-PAH (EW7.7\,$>$\,0.9\um) sources where it is clear that the bulk of these sources have at least one far-IR detection. The middle column shows weak-PAH sources (EW7.7\,$<$\,0.9\um), where the SED fitting suggests star-formation dominates (see Section\,\ref{sec_lir}), a conclusion that is clearly the result of the bulk of these sources having at least one far-IR detection. Lastly, the right-hand column shows sources that are both AGN-dominated in the mid-IR (i.e. weak-PAH), and where the SED fitting suggests  the AGN contributes to $>$\,50\%\ of $L_{\rm{IR}}$ (see Section\,\ref{sec_lir}). Roughly half of these sources are without detections in the far-IR.

\subsection{IR luminosities and AGN fractions \label{sec_lir}}

We derive the infrared luminosities ($L_{\rm{IR}}$) of our sources as the integral over the 3\,--\,1000\um\ range of the best-fit SED model for each source. The uncertainty on $L_{\rm{IR}}$ derives from the MCMC fitting and represents the 68\%\ confidence level (see Appendix\,\ref{sec_mcmc} for details). The "hot dust" component of our composite model is qualitatively similar to the torus models as seen in Figure\,\ref{fig_agnmodels}. We therefore define $L_{\rm{AGN}}$ as simply the integral of the hot component over the 3\,--\,1000\um\ range as was done for $L_{\rm{IR}}$.  The AGN fraction is then defined as $L_{\rm{AGN}}/L_{\rm{IR}}$.  It is clear from Figure\,\ref{fig_seds} that the interpretation of these SED fits is not only complicated by the lack of far-IR detections in some sources, but also by the model assumptions. For example, a model that effectively adopts a more compact torus whose emission peaks at shorter wavelengths \citep[such as][]{polletta08} would lead to the far-IR detections in MIPS8242 being ascribed to star-formation, instead of to the RJ tail of the hot/torus component as here. We address the uncertainties in our SED fits in Appendix\,\ref{sec_mcmc}. 

In order to allow for direct comparison with other samples as well as various diagnostics, we also compute a number of rest-frame monochromatic luminosities, specifically at 5.8\um, 8.0\um, 15\um, and 30\um. The 8.0\um\ is always covered by the IRS spectrum, and hence it is computed directly from that spectrum (with the IRAC8um filter overlaid). The rest-frame 5.8\um\ and 15\um\ luminosities are also determined from the IRS spectrum whenever possible or from the SED fits otherwise. The 30\um\ luminosity is always measured from the SED fits. In all cases, square filters with $\Delta\lambda$/$\lambda$=0.033 are used in order to allow for direct comparison with \citet{veilleux09}. The $L_{\rm{IR}}$ and $L_{\rm{AGN}}$ values as well as these monochromatic luminosities are all given in Table\,\ref{table_lums}. 

Lastly, we note that many of the AGN luminosities we derive, including all $z$\,$>$\,1 AGN-dominated, or composite systems are $>$\,$10^{12}$\lsun. This places our sources in the quasar regime. On the other hand, by selection, Type-1 quasars are excluded from our sample, therefore we are seeing obscured quasars. 

\subsection{Star-formation rates \label{sec_sfr}}

We can estimate the starburst luminosity simply by $L_{\rm{SB}}$\,=\,$L_{\rm{IR}}$-$L_{\rm{AGN}}$. We convert $L_{\rm{SB}}$ to star-formation rate (SFR) using the relation in  \citet{kennicutt98}. A small caveat here is that while we used the integrated 3-1000\um\ emission for $L_{\rm{IR}}$ and $L_{\rm{AGN}}$ and hence $L_{\rm{SB}}$, in \citet{kennicutt98} the 8-1000\um\ range is used. The difference between the two is negligible for the SEDs of star-forming galaxies, especially compared with the uncertainties in in the relative AGN/SB contribution here.  A more serious caveat is that the stellar initial mass function (which underlies all such conversion relations) is unknown for our sources. Therefore, these SFR values assume that the basic stellar population and dust properties are the same for our sources as in normal star-forming galaxies nearby. For example, a more top-heavy IMF, as suggested by some theoretical models \citep{baugh05}, would result in  smaller SFR values. We have no means of addressing these systematic uncertainties here, therefore simply apply the most commonly used $L_{\rm{IR}}$-SFR conversion relation in the literature. The derived SFR values are listed in Table\,\ref{table_lums}.  These are typically in the range $\sim$\,100\,--\,500\msun/yr for the $z$\,$<$\,1.5 starburst or composite sources and $\sim$\,2000\msun/yr for the $z$\,$>$\,1.5 starburst sources, and in-between the above extremes ($\sim$\,500\,--\,1500\msun/yr) for the higher-$z$ composite sources. As stated in the introduction, the small number of sources for which we have CO measurements show significant masses of cold molecular gas -- sufficient to fuel such extreme levels of star-formation \citep{yan10}. 

\begin{figure*}[t!]
\epsfig{file=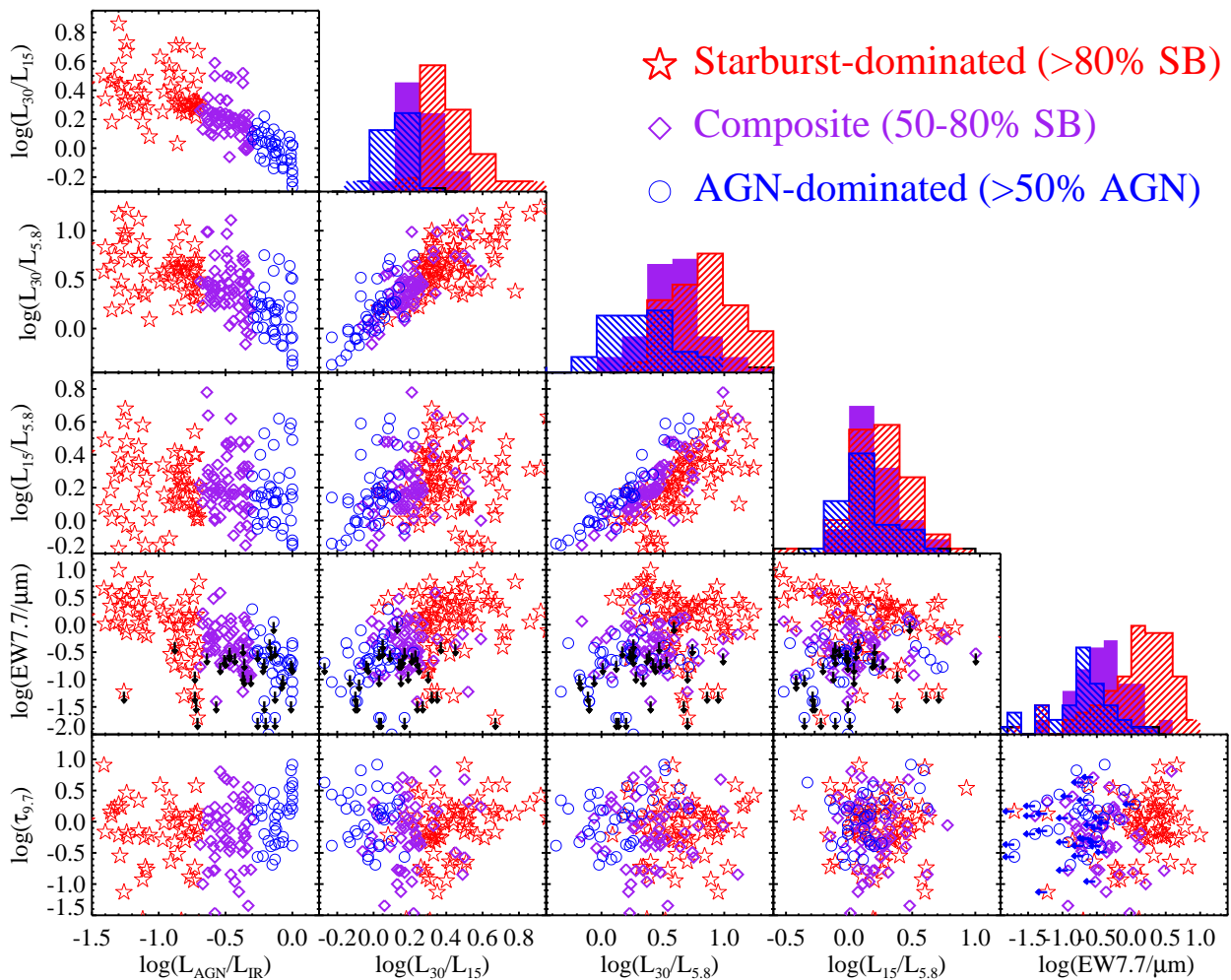, clip=, width=1.0\linewidth} 
\caption{A variety of possible starburst-AGN diagnostics. Here, the red stars are the "starbursts" with $<$\,20\% AGN, the purple diamonds are the "composites" with 20-50\% AGN, while the blue circles are the "AGN" with $>$\,50\% AGN fraction. The 2\,$\sigma$ upper limits on the 7.7\um\ equivalent widths are also indicated. \label{fig_diag}}
\end{figure*}

\subsection{Mid-IR SB-AGN diagnostics \label{sec_diag}}
Figure\,\ref{fig_diag} show a mosaic of different mid-IR AGN-starbursts diagnostics \citep{veilleux09}, where in particular we look for trends between the mid-IR properties of galaxies and their overall IR SED properties, in particular the overall fraction AGN to the total infrared power (i.e. $L_{\rm{AGN}}/L_{\rm{IR}}$). To help emphasize the trends with relative AGN power, we also divide the sample in three categories: sources with AGN-fraction of $<$\,20\% are "starburst-dominated"; sources with AGN-fractions above 50\% are AGN-dominated; and sources in between these limits are starburst-AGN composites. Following the discussion in Appendix\,\ref{sec_mcmc}, it should be kept in mind that the boundaries especially between the composites, and starbursts on one side and the AGN on the other are blurred.  

\begin{figure*}[t!]
\centering
\epsfig{file=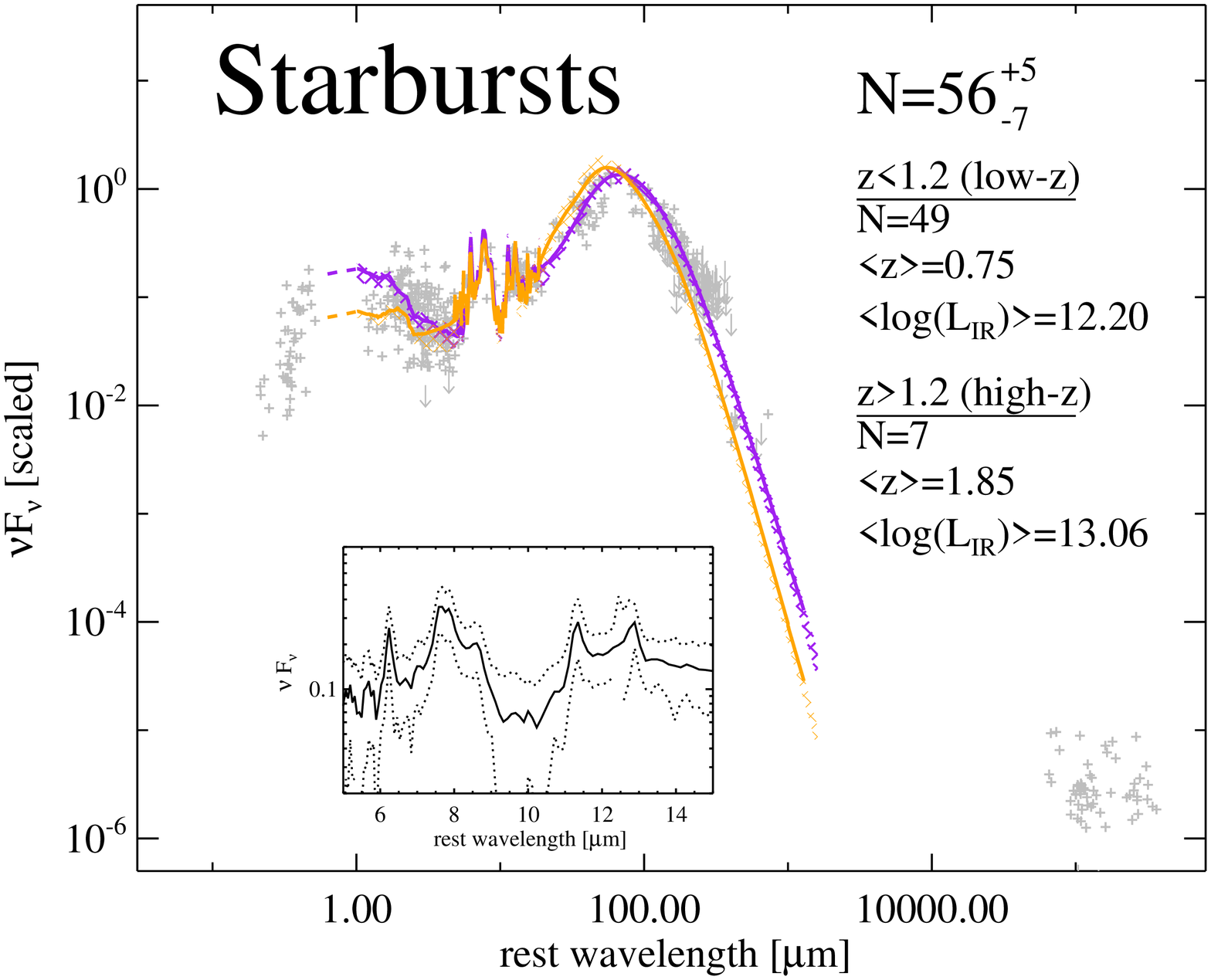, clip=, width=0.32\linewidth} 
\epsfig{file=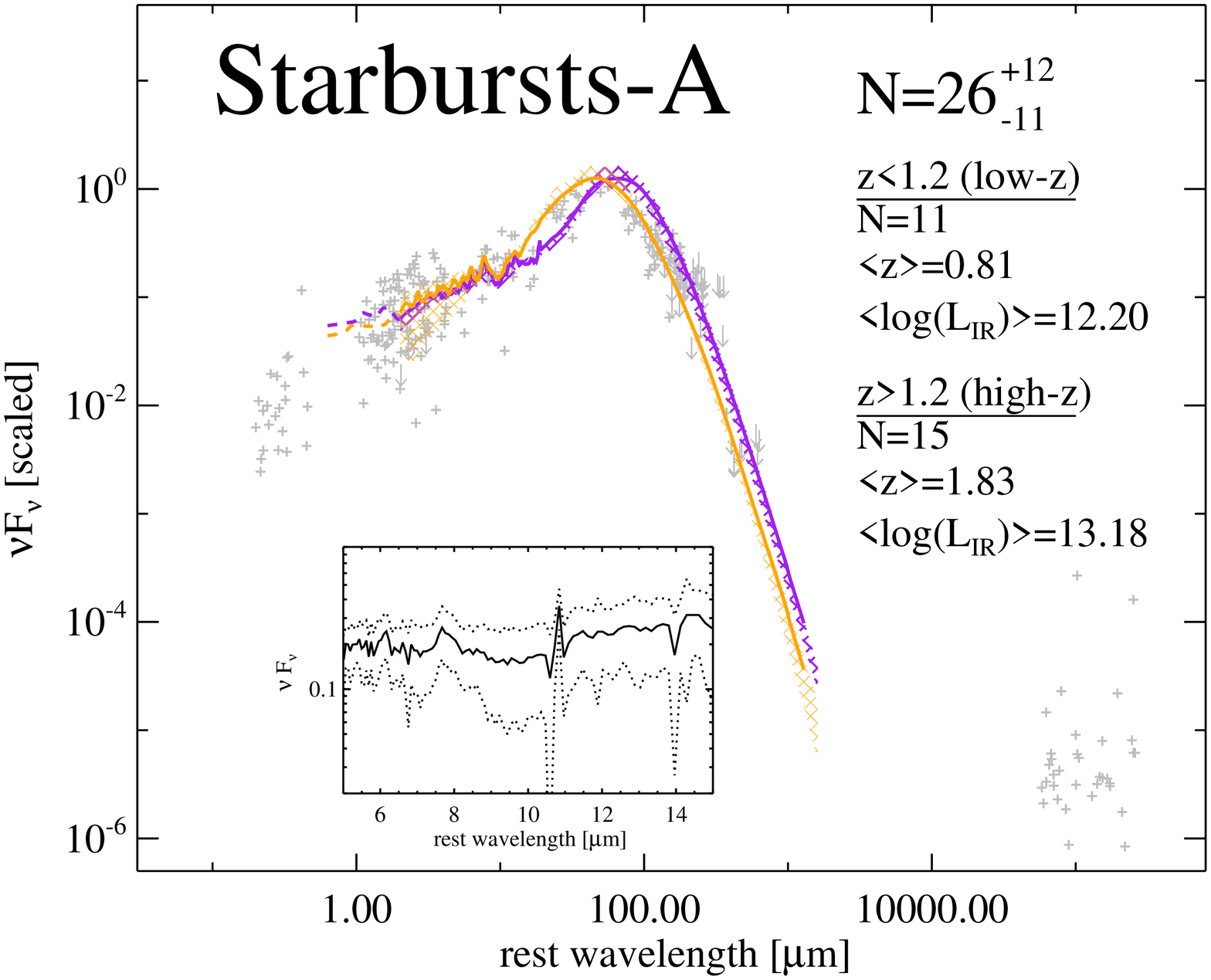, clip=, width=0.32\linewidth} 
\epsfig{file=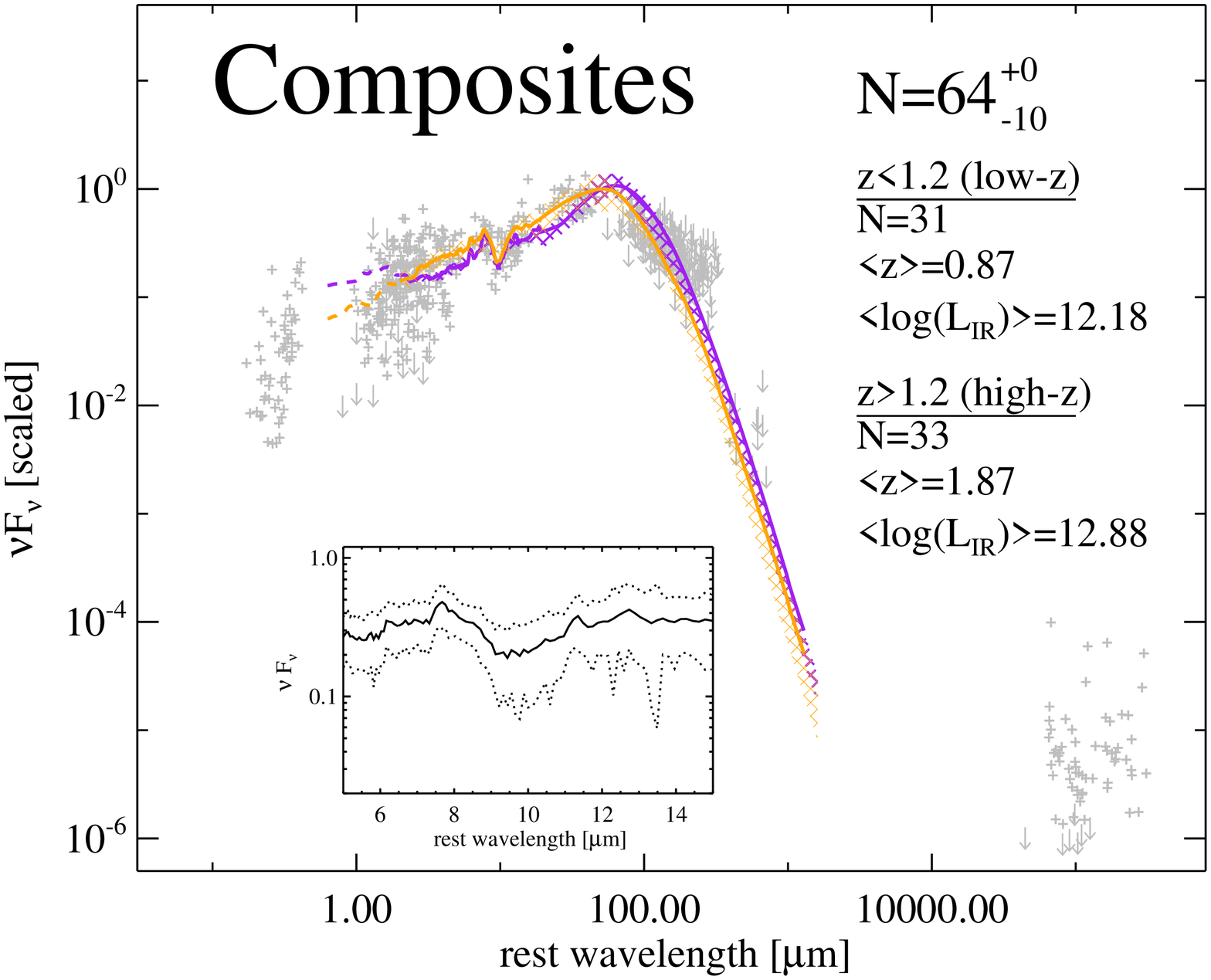, clip=, width=0.32\linewidth}\\
\epsfig{file=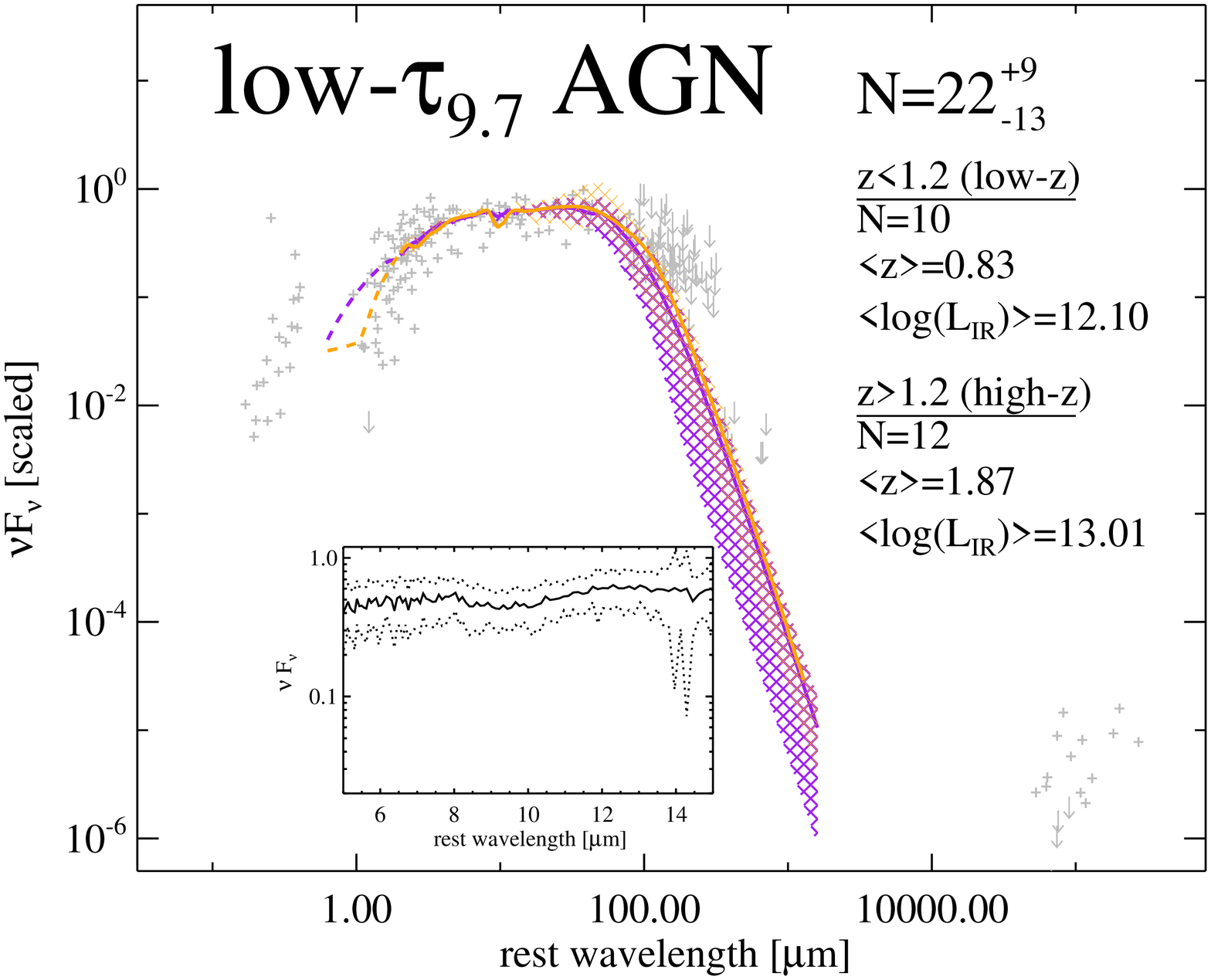,clip=, width=0.32\linewidth}
\epsfig{file=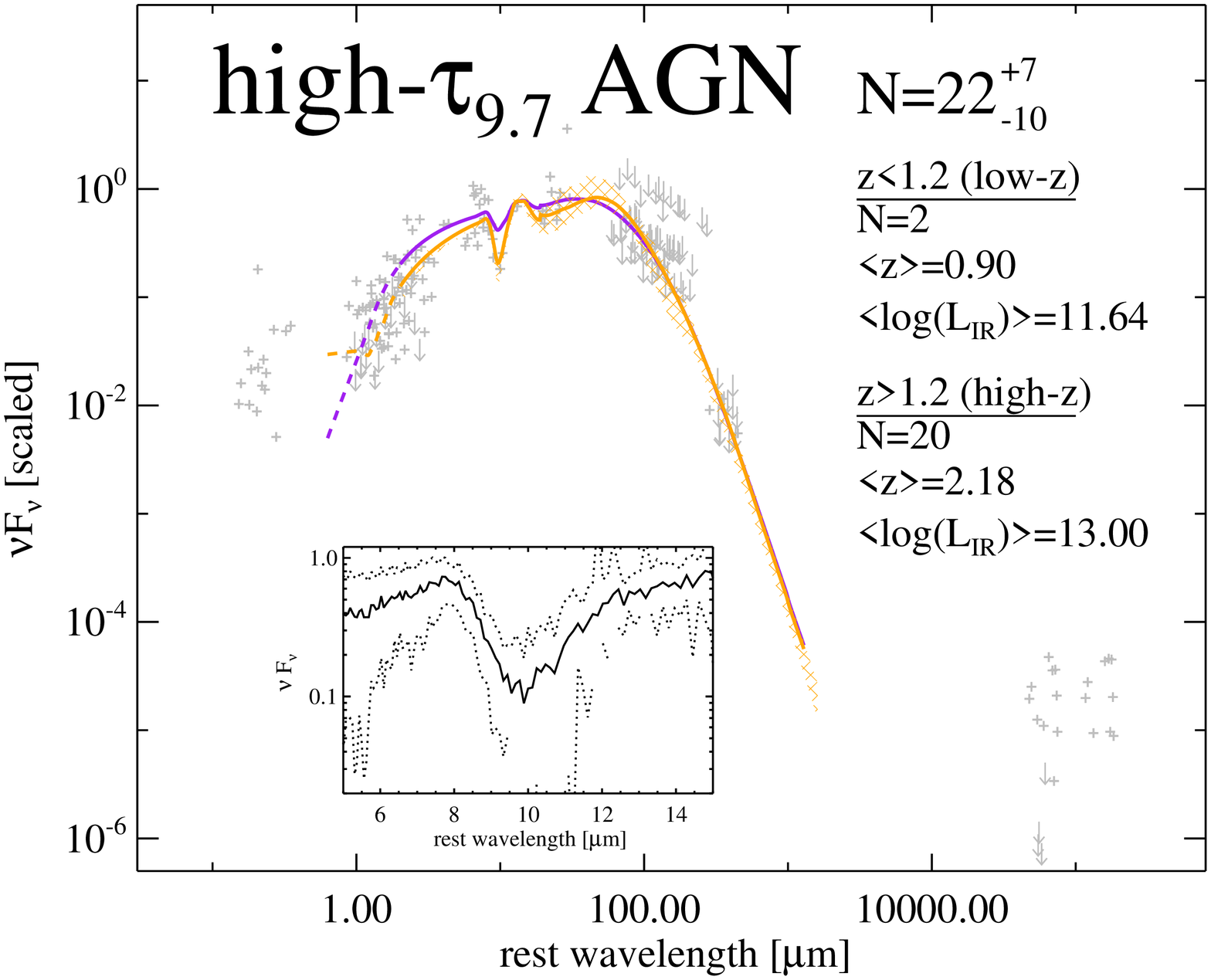,clip=, width=0.32\linewidth}
\caption{Median SEDs constructed for different classes of sources (see text for details) and for two redshift bins: $z$$\leq$1.2 and $z$\,$>$\,1.2. For each redshift bin, we indicate the number of sources, and the mean redshift and luminosities. Individual source SEDs are scaled to each source's total power output. Upper limits are 3\,$\sigma$. The uncertainties on the total number of sources in each category, as well as the spread in these median templates derive from the spread in individual source SED's when all solutions within $\chi^2$\,$<$\,$\chi^2_{min}+1$ are considered.  The parts of the SED templates marked with dashed lines are considered highly uncertain. Note that since there are only 2 high-$\tau$ AGN sources in the low-$z$ bin, the uncertainty for that template could not be determined.  In all cases, the purple (darker) curves represent the low-$z$ templates while the orange (lighter) curves represent the high-$z$ templates. \label{fig_templates}}
\end{figure*}

From Figure\,\ref{fig_diag} it is clear that the $\log(L_{30}/L_{15})$ color and to a slightly lesser extend the $\log(L_{30}/L_{5.8})$ color and  the PAH equivalent width are all reasonably good tracers of the overall AGN-fraction, though with substantial scatter. For example, the scatter in $\log(L_{30}/L_{5.8})$ among the AGN-dominated sources is likely due to the fact that some AGN-dominated sources can be very red in this color due to steeper $\alpha$ and heavy obscuration (e.g. MIPS8242, Figure\,\ref{fig_seds}). Only one of the AGN-dominated sources,  MIPS277, is a strong-PAH source (EW7.7\,$>$\,0.9\um), however this source is borderline with an AGN fraction $\sim$\,50\%. We note that about 1/3 of the starburst-dominated sources are weak-PAH (EW7.7\,$<$\,0.9\um). We examined these sources and found in nearly all cases, strong far-IR detections therefore their classification as starbursts is likely correct, despite them being dominated by AGN in the mid-IR regime. We return to this in the following Section\,\ref{sec_templates}.

The $\log(L_{15}/L_{5.8})$ and $\tau_{9.7}$ parameters do not trace the AGN-fraction. Even excluding the starburst dominated sources, the large scatter in both parameters indicates a wide range in mid-IR spectral shapes (likely related to obscuration levels) among our AGN and composite sources. This is the result of a lack of color-selection in this sample, in contrast to our earlier studies of the GO1 sample \citet{yan07,sajina07a} where the sources were found to be both redder and with on average deeper silicate absorption than seen here.  Specifically, the mid-IR spectral indices of the AGN-dominated sources here are typically $\alpha$\,$\sim$\,1\,--\,1.5, similar to local AGN and quasars \citep{netzer07,mullaney11}.  This is in contrast to our earlier findings for the GO1 sample alone, where the typical spectral index was $\alpha$\,$\sim$\,2 \citep{sajina07a}, due tot he additional color-selection in that sample. Unsurprisingly, the bulk of our high-$\tau_{9.7}$ AGN-dominated sources, come from the GO1 sub-sample. 

\begin{figure*}[t!]
\epsfig{file=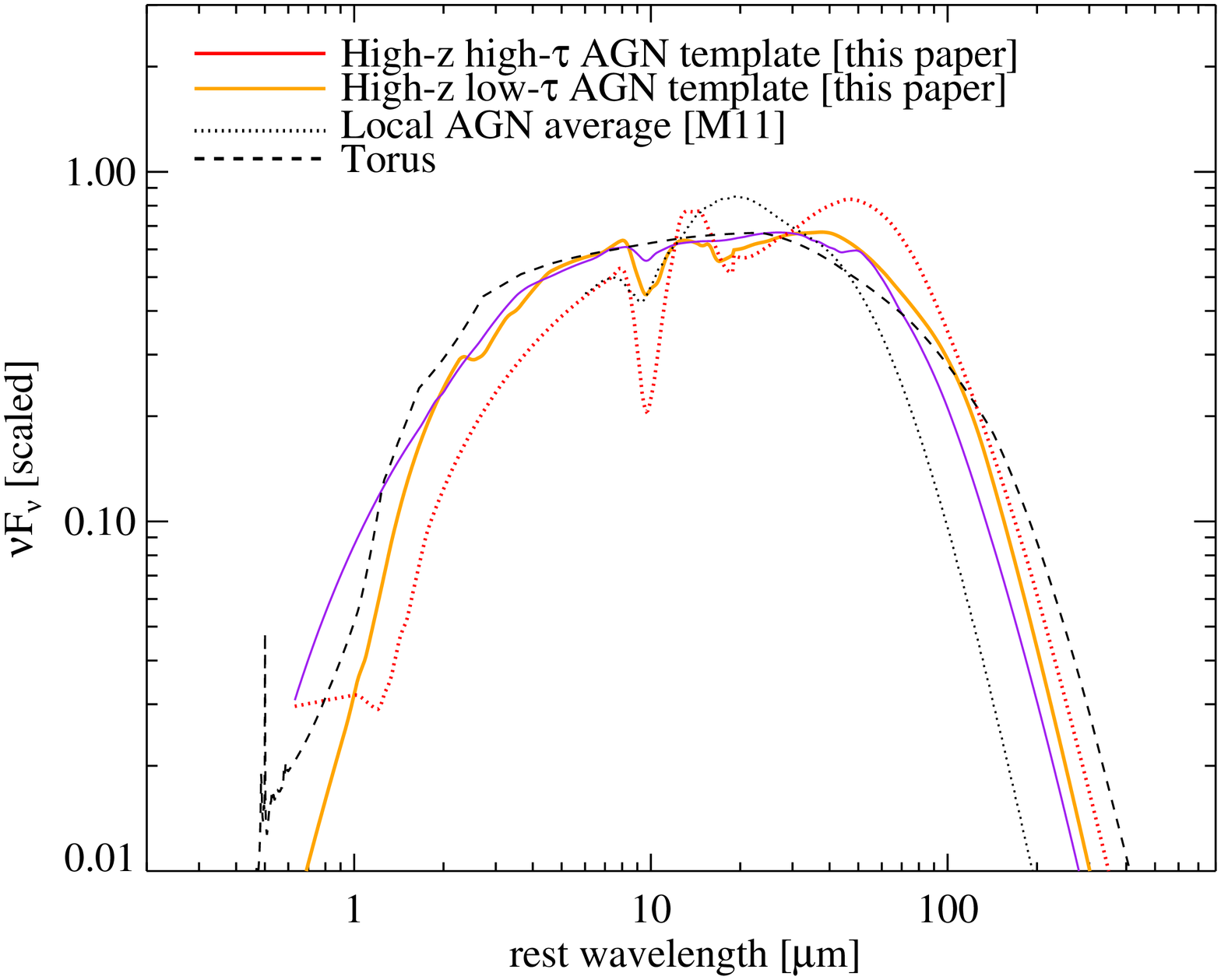, clip=, width=0.45\linewidth}
\epsfig{file=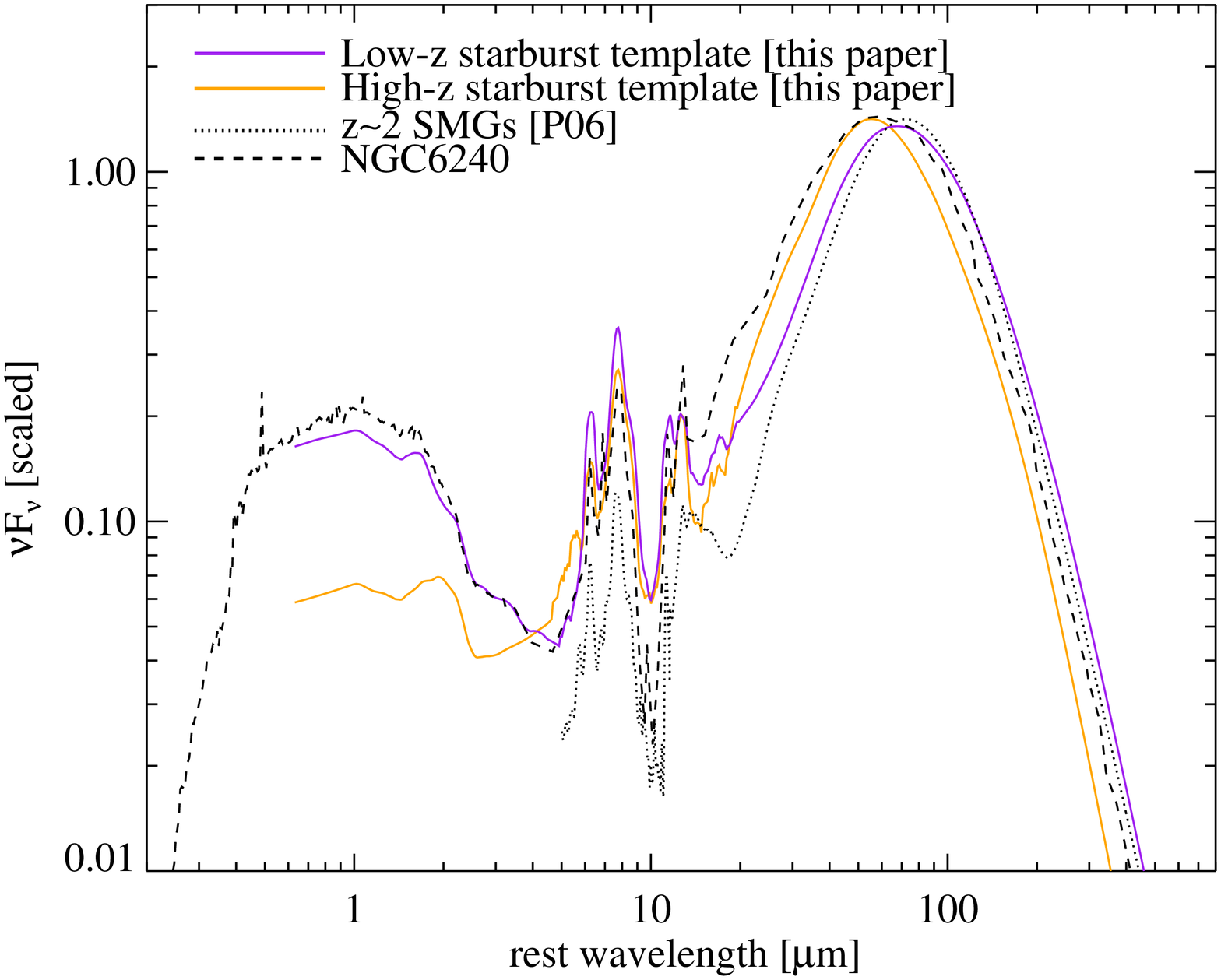, clip=, width=0.45\linewidth} \\
\epsfig{file=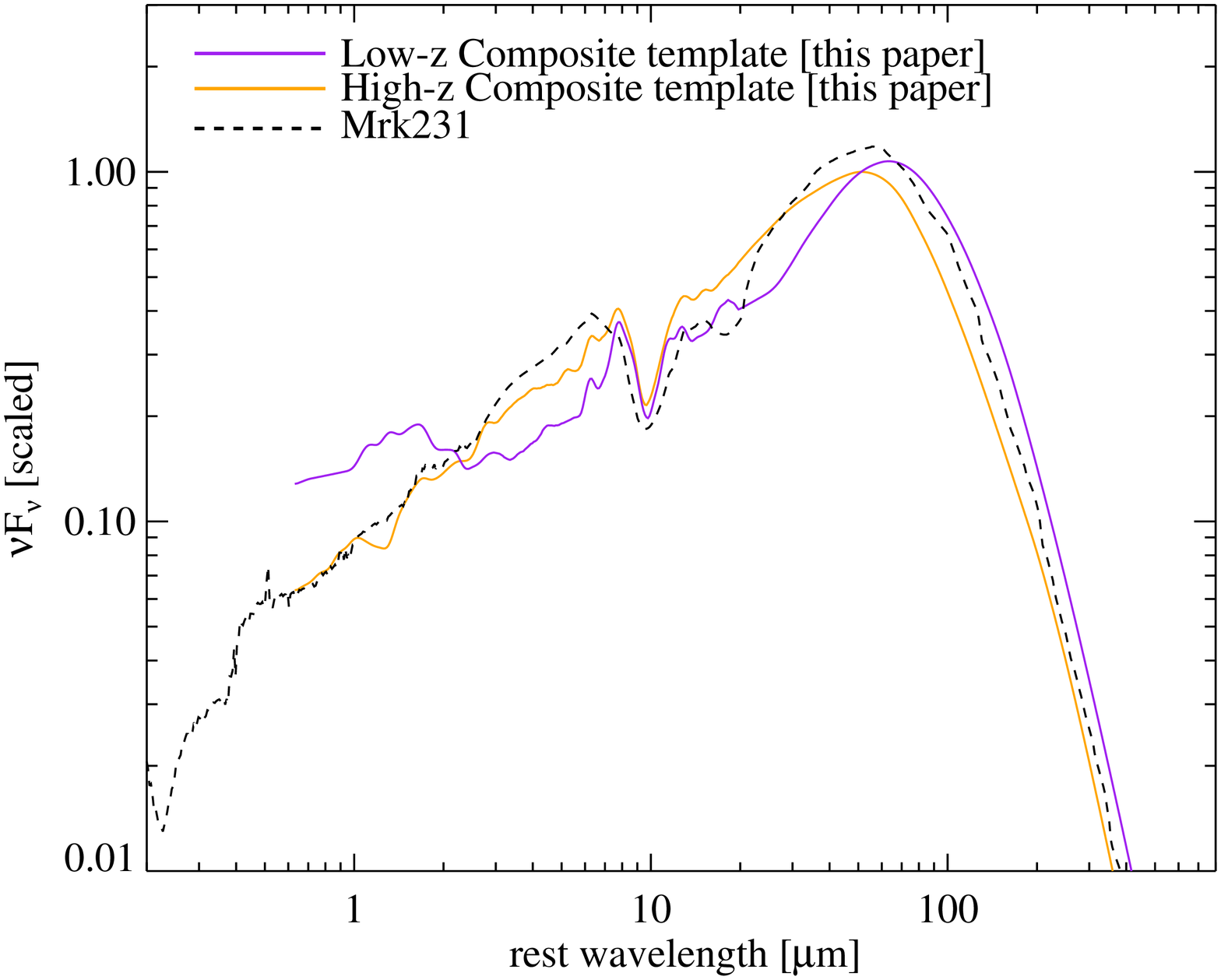, clip=, width=0.45\linewidth}
\epsfig{file=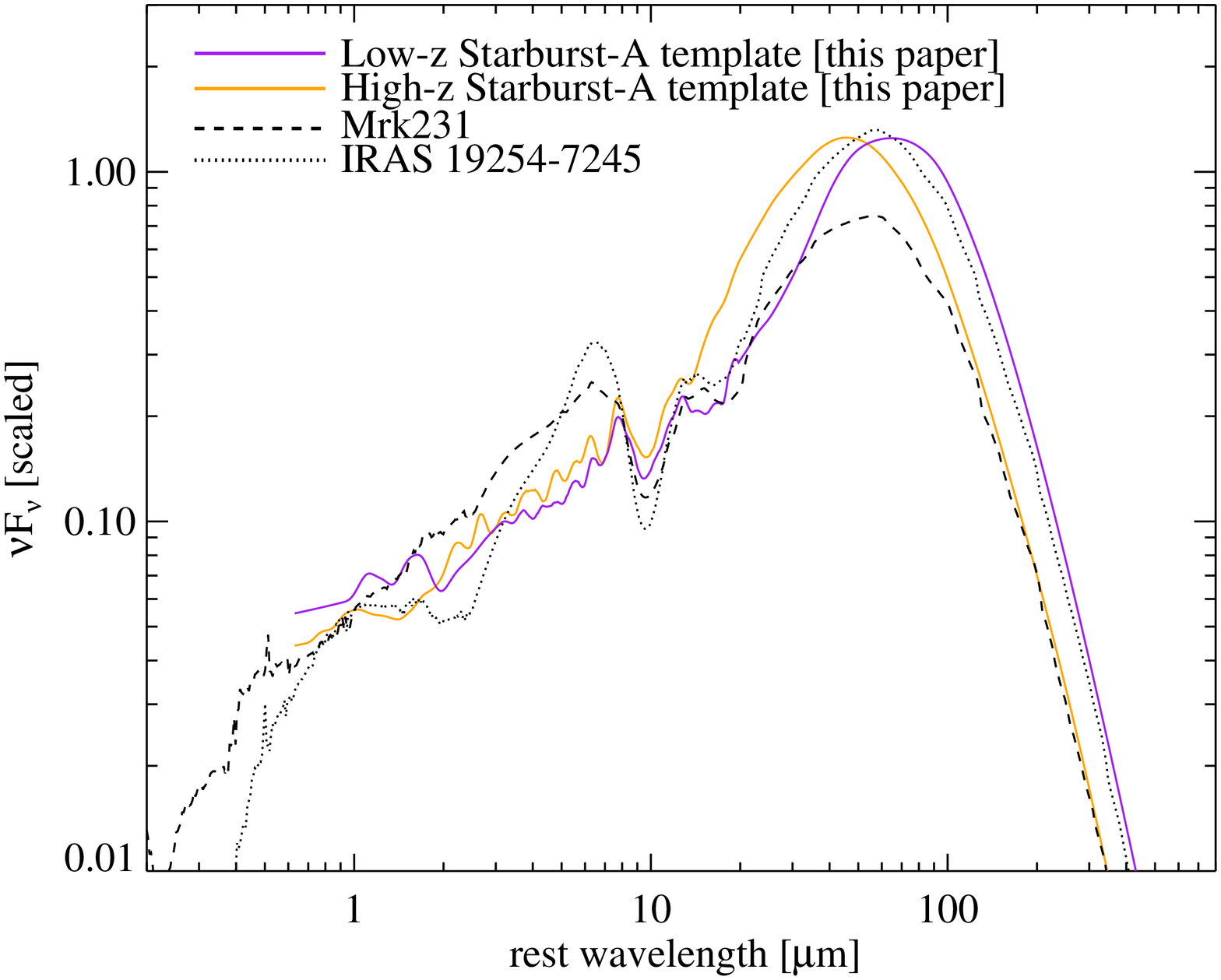, clip=, width=0.45\linewidth} 
\caption{Our median SEDs compared with other starburst or quasar templates from the literature. The local AGN average template is from \citet{mullaney11}. The SMG template is from \citet{pope06}. The rest of the templates are from the SWIRE template library \citep{polletta08}. Mrk231 and IRAS 19254-7245 are both Seyfert 2 ULIRGs. \label{fig_templates_compare}}
\end{figure*}

Lastly, the finding that  $\log(L_{30}/L_{15})$ is a reasonable tracer of the overall AGN fraction is consistent with the same finding for local ULIRGs in \citet{veilleux09}. However, even our starburst sources are consistently bluer than the local ULIRGs. This may be indicative of higher AGN contributions in our sources, or SED evolution, as discussed in Section\,\ref{sec:sedevol}.

\subsection{Templates for different SED types \label{sec_templates}}

It is clear that the SEDs of our galaxies have a wide range of properties (see example Figure\,\ref{fig_diag}). Here we would like to effectively summarize these properties, by dividing the sample into a few categories of source SEDs. To determine the classes, we both look at the overall AGN/starburst fractions (i.e. the far-IR properties) and the mid-IR properties. The latter are based on the PAH strength as parameterized by the 7.7\um\ equivalent width (EW7.7) as well as the silicate feature depth as parameterized by $\tau_{9.7}$ (see Section\,\ref{sec_mir} for a description of how these quantities are derived). The reason why we find it useful to look at both the far-IR and mid-IR SED for this classification is that we find the mid-IR classification does not map one-to-on onto a classification based on the overall IR SED, and vice versa. For example, sources classified as AGN-dominated in the mid-IR often are seen as starburst-dominated when the far-IR data are included.  Our classification is based on the following categories:

\begin{itemize}
\item low-$\tau_{9.7}$ AGN: sources with AGN fraction of $\geq$\,50\% and with $\tau_{9.7}$\,$<$\,1 
\item high-$\tau_{9.7}$ AGN: sources with AGN fraction of $\geq$\,50\% and with $\tau_{9.7}$\,$\geq$1
\item Composites: sources with AGN fractions in the range 20\,--\,50\%
\item Starbursts-A: sources with AGN fraction of $<$\,20\%, but with significant mid-IR AGN as indicated by EW$_{7.7}$\,$<$\,0.9\um\
\item Starbursts: sources with AGN fraction of $<$\,20\% and with EW$_{7.7}$\,$\geq$\,0.9\um\
\end{itemize}

In Figure\,\ref{fig_templates}, we show the median templates in each category, constructed from the best-fit individual source SEDs scaled by their total power output. To give a sense of the spread about these templates, we also show the rest-frame broadband data for each source with the same scaling factor applied. The insets in each of the panels in Figure\,\ref{fig_templates} show the average mid-IR spectrum along with its 1\,$\sigma$ spread. For example, by selection, the "Starbursts" show strong PAH features in their average spectrum, where the 6.2, 7.7, 8.6, 11.3, and 12.6\um\ features are all clearly visible, whereas the "Starburst-A" sources show weak or no PAH features in the mid-IR combined with a far-IR peak. The composite sources also show a hint of PAH features in that the 7.7, 11.3, and 12.6\um\ features can be discerned, though much more weakly than for the "Starburst" sources. The two AGN templates are best described as continuum spectra with the only feature being due to silicates absorption.  These also typically have upper limits in the far-IR that indeed preclude the significant presence of a cold dust far-IR peak. These average templates derived from the IRS supersample are available online\footnote{\url{http://cosmos2.phy.tufts.edu/~asajina/IRSsupersample.html}}.

In Figure\,\ref{fig_templates_compare}, we compare the above templates with a few illustrative local source templates. The low-$\tau_{9.7}$ AGN template is reasonably similar to the local AGN template of \citet{mullaney11} and generally agrees with classic torus models such as the Torus template in the SWIRE library. The high-$\tau_{9.7}$ AGN template is distinctly redder and even shows somewhat stronger emission past $\sim$\,20\um. Such sources differ from local AGN sources which tend not show such extreme silicate absorption features (see Section\,\ref{sec:miragn}).  The starburst templates look like fairly standard starburst galaxy templates. Here we compare them with the sub-mm galaxy template of \citet{pope06} showing that our sources have relatively higher mid-IR emission (as expected given our selection), but comparable dust temperatures to that of SMGs. When available, their sub-mm fluxes suggest that they would meet the criteria for being SMGs \citep{sajina08}.  We also compare with SED of NGC6240, a well studied local starburst galaxy with luminosity comparable to our lower-$z$ sample.  Its SED agrees well with our starbursts although showing somewhat higher dust temperature. We discuss in more detail how our starburst sources compare with local LIRGs and ULIRGs in Section\,\ref{sec:sedevol}. The composite and starburst-A sources have rather similar SEDs, except that the latter has a stronger far-IR emission. To illustrate this, and highlight their composite nature, we compare both to Mrk231, the quintessential AGN-starburst composite in the local universe. 

\begin{figure}[h!]
\epsfig{file=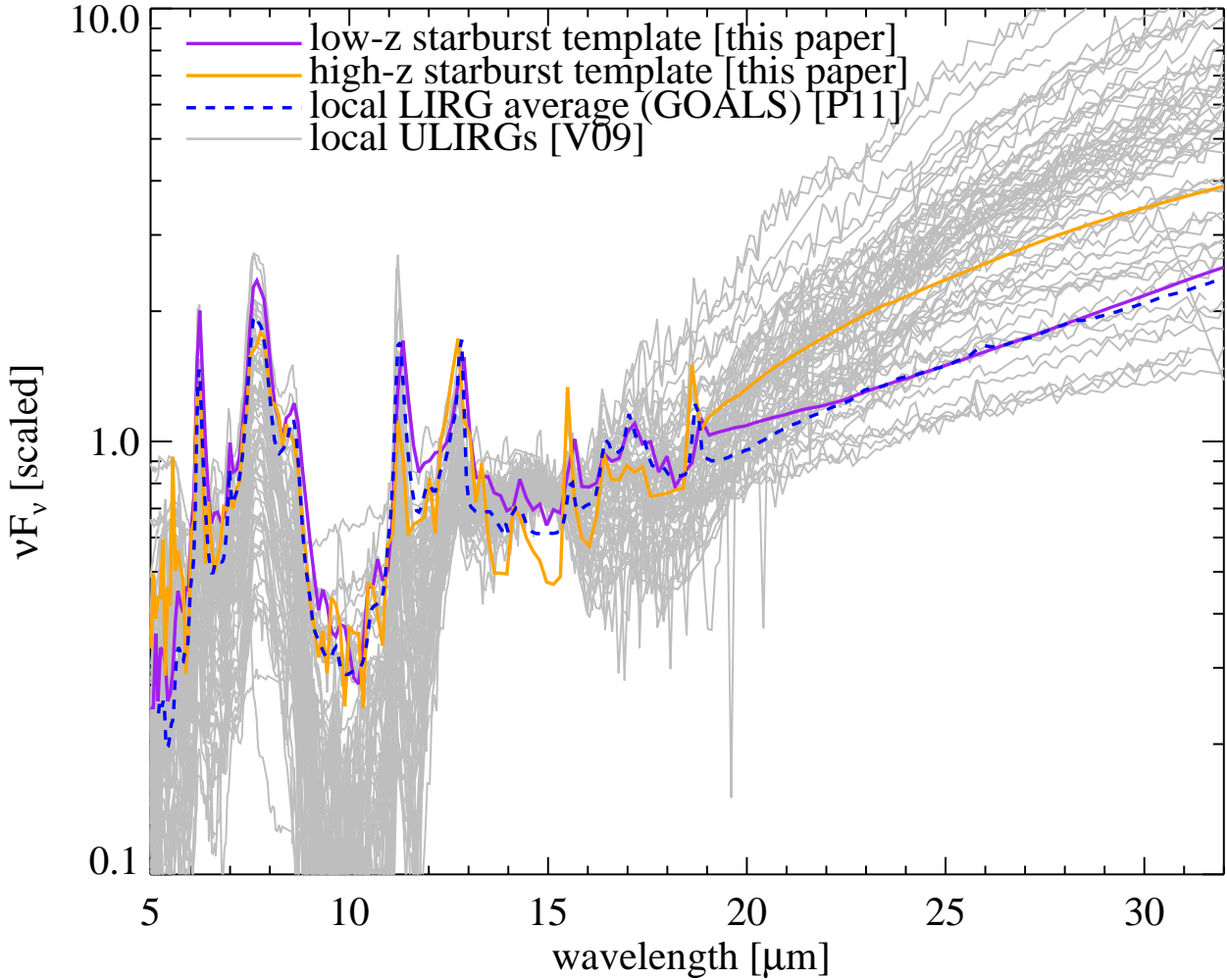,clip=,width=1.0\linewidth}
\caption{The average mid-IR SEDs of our low-$z$ and high-$z$ starburst-dominated sources, compared with the mid-IR spectra of  local ULIRGs (thin grey curves) and the average local LIRGs spectrum of \citet{petric11}. \label{fig_mir_sed}}
\end{figure}

\begin{figure}[h!]
\epsfig{file=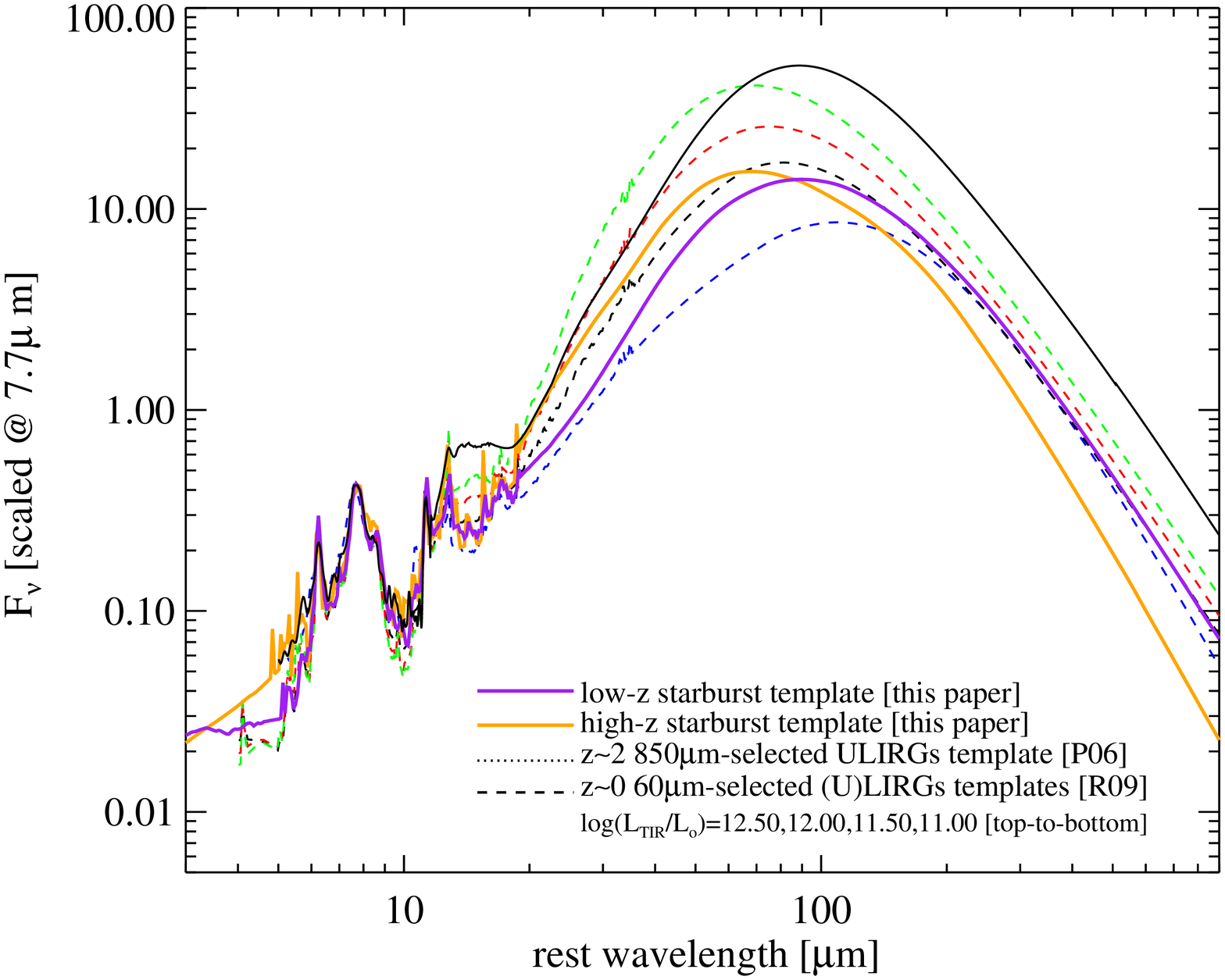,clip=,width=0.5\linewidth}
\epsfig{file=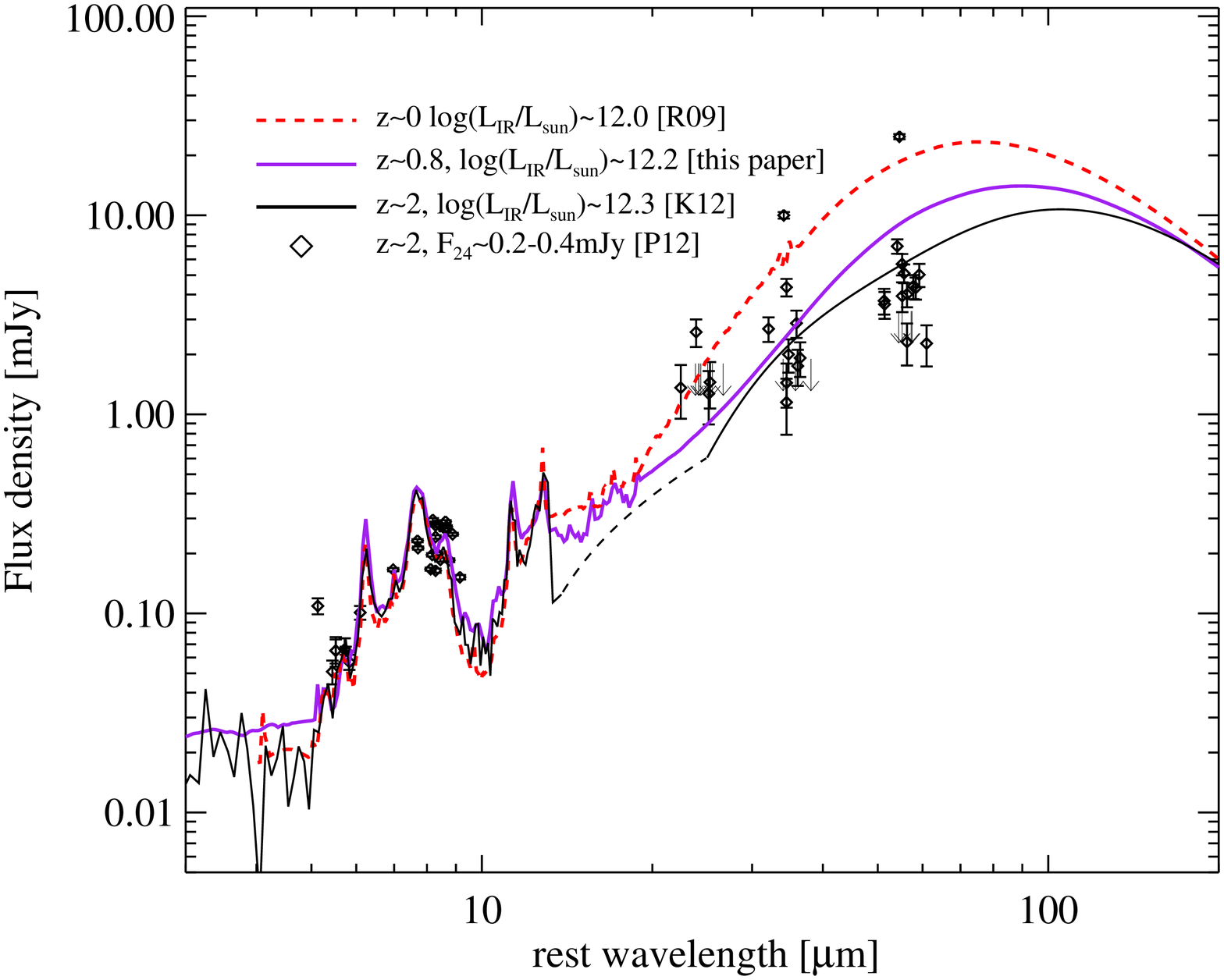,clip=,width=0.5\linewidth}
\caption{A comparison of our starburst templates with other templates of comparable luminosity and/or redshift in the literature. We scale all at 7.7\um\  which emphasizes the spread in the far-IR. {\it Left:} Here we plot SEDs for local IR-luminous star-forming galaxies \citep[][here R09]{rieke09}, emphasizing the well known evolution of IR SEDs with luminosity.  We then overplot our $z$\,$\sim$\,0.8 (low-$z$) and $z$\,$\sim$\,1.8 (high-$z$) starburst templates, which clearly are much closer to local sources about an order of magnitude less luminous than themselves, than to sources of comparable luminosity (as already seen in Figure\,\ref{fig_mir_sed}). This however is a function of selection since sub-mm galaxies (SMGs) of comparable luminosity and redshift to our higher-$z$ sources show much higher far-IR to mid-IR ratios. {\it Right:}  Here we focus on the evolution of a $\sim$\,$10^{12}$\lsun\ galaxy SED from $z$\,$\sim$\,0 to $z$\,$\sim$\,2. For the higher-$z$ template, we use the SED composite of Kirkpatrick et al. (2012, in press; here K12). We overplot the broadband data for the $z$\,$\sim$\,2 $F_{24}$\,=\,0.2\,--\,0.3\,mJy $z$\,$\sim$\,2 and $L_{\rm{IR}}\sim10^{12}$\lsun\ sources from \citet{pozzi12}, which are consistent with the K12 template. This figure suggests that the SEDs of IR-luminous galaxies not only evolve strongly, but also show a big spread for a given luminosity and redshift (our high-$z$ starbursts have comparable luminosities and redshifts to the SMGs, but very different far-IR to mid-IR ratios).  \label{fig_compare_templates}}
\end{figure}

While we find these templates useful, in the joint analysis of the mid-IR and far-IR properties of our sample, it is clear that for example the "Starbursts-A" and "Composites" are both really sources that are dominated by star-formation in the far-IR, but by AGN in the mid-IR. We can therefore summarize these classes as follows.  A total of 146 sources (76\% of the sample) have $<$\,50\% of their $L_{\rm{IR}}$ contributed to by an AGN.  However, of these, 90 (the composite and Starburst-A sources), are dominated by AGN in the mid-IR as indicated by a low PAH 7.7\um\ feature equivalent width. A total of 45 sources\footnote{In the plots in Figure\,\ref{fig_templates}, MIPS277 is excluded as it is borderline AGN with strong PAH.} have $<$\,50\% of their $L_{\rm{IR}}$ contributed to by an AGN. Therefore, the essentially pure starbursts are $\sim$\,30\%\ (nearly all at $z$\,$<$\,1.2), the pure AGN are $\sim$\,23\%, and the composites are $\sim$\,47\%\ of the total sample. The uncertainties on these fractions are somewhere between 5 and 10\%\ (see Section\,\ref{sec_mcmc}). 

\section{Discussion}

\subsection{Evidence for SED evolution \label{sec:sedevol}}

For our purposes, "SED evolution" means that the typical SED of a source of a given IR luminosity at higher redshifts differs from the typical SED of a source of the same luminosity at redshift $\sim$\,0. An important caveat is that given our mid-IR selection, our sample is not necessarily representative of all galaxies at a given luminosity and redshift. However, evidence for SED evolution has been shown earlier for far-IR selected sources \citep{huynh07_smg,pope08,seymour10}. The combination of these earlier results and our results here do suggest that the typical IR-bright galaxy at high redshift is indeed different from the typical IR-bright galaxy locally.
 
The bulk of our sources are ULIRGs ($L_{\rm{IR}}$\,$\ge$\,$10^{12}$\lsun), with $\sim$\,10\% of our sample (the lowest-$z$ sources) being Luminous Infrared Galaxies (LIRGs; $L_{\rm{IR}}$\,=\,$10^{11}$\,--\,$10^{12}$\lsun).  In Figure\,\ref{fig_mir_sed}, we show the average low-$z$ and high-$z$ templates derived for our starburst-dominated sources (see Section\,\ref{sec_templates}) compared with local LIRGs and ULIRGs. The local LIRG comparison sample is the Great Observatories All-sky LIRG Survey \citep[GOALS;][]{armus09}, which represent a complete sub-set of the {\sl IRAS} 60\um-selected local LIRGs. Here we specifically make use of the average GOALS LIRG mid-IR spectrum as computed by \citet{petric11}.  For our local ULIRG   comparison sample, we use the 1\,Jy local ULIRGs sample which represents all 118 {\sl IRAS} $S_{60\mu m}$\,$>$\,1\,Jy ULIRGs within $z$\,$\sim$\,0.3 from the redshift survey of \citet{ks98}. Specifically, we make use of all available mid-IR IRS spectra (74) of these 118 sources as presented in \citet{veilleux09}.  Therefore, both our local LIRG and ULIRG comparison samples are ultimately based on a 60\um\ flux-density selection. 

The most striking conclusion from Figure\,\ref{fig_mir_sed} is that while some local ULIRGs have colors comparable to our sample, we find better agreement with the average LIRG spectrum.  This result is particularly surprising for the higher-$z$ starburst dominated sources which have an average luminosity of $\sim$\,7\,$\times$\,$10^{12}$\lsun. The typical local ULIRG is distinctly redder in the 15\,-\,to\,-\,30\um\ regime, as well as having deeper silicate absorption features at both 9.7 and 18\um, implying high levels of obscuration. Indeed, most ULIRGs are associated with late stage mergers when the overall level of obscuration is maximal. 

In Figure\,\ref{fig_compare_templates}, we extend the mid-IR spectral comparison of Figure\,\ref{fig_mir_sed} to the entire IR SED. Here we compare our starburst templates with the \citet{rieke09} templates -- the most up to date local galaxy-derived templates that characterize the IR SEDs of starburst galaxies split into logarithmic bins in IR luminosity. The galaxy sample behind these templates is ultimately again the {\sl IRAS} 60\um\ selected LIRGs and ULIRGs. However, \citet{rieke09} trim these significantly by the requirement that the starburst nature of these sources be firmly established in the literature through ancillary data, as well as that they have good spectral coverage across the IR SED. As discussed earlier, in local galaxies, as the IR luminosity increases the ratio of mid-IR to far-IR luminosity decreases, the 15-30\um\ color reddens, and the silicate feature deepens. All of these are likely indicative of increasing overall obscuration as we progress from the lowest luminosity LIRGs ($L_{\rm{IR}}$\,$\sim$\,$10^{11}$\lsun) to the highest luminosity ULIRGs ($L_{\rm{IR}}$\,$\gs$\,6\,$\times$\,$10^{12}$\lsun). Both our lower-$z$ and higher-$z$ starburst-dominated sources show SEDs that correspond to local sources in the range $\log(L_{\rm{IR}})$\,$\sim$\,11.25-11.50. However, the mean luminosities for the low-$z$ and high-$z$ samples are respectively $\sim$\,$10^{12.2}$\lsun\ and  $\sim$\,$10^{13}$\lsun, with the mean redshifts being respectively 0.8 and 1.9. Therefore, it is obvious that our sample shows strong SED evolution, with the best local analogs to our high-$z$ sources being sources of significantly lower overall power output. It is worth noting that this discrepancy is even more dramatic than the difference observed between local ULIRGs and SMGs the latter having colder dust temperatures, but otherwise comparable far-IR to mid-IR ratios to those of local ULIRGs. This is a reflection of our 24\um\ selection which biases us toward stronger mid-IR emission. 

Lastly, we want to address whether or not this strong SED evolution is also present in fainter mid-IR selected sources (i.e. reaching lower luminosities for a given redshift). Here we look at the $F_{24}$\,=\,0.2\,--\,0.3\,mJy $z$\,$\sim$\,2 sources from \citet{pozzi12}. These sources have $z$\,$\sim$\,2 and $L_{IR}$\,$\sim$\,$10^{12}$\lsun. The right-hand panel of Figure\,\ref{fig_compare_templates} shows that the typical far-IR to mid-IR ratios of these fainter sources are very far from the expectations of the local $L_{IR}$\,$\sim$\,$10^{12}$\lsun\ template \citep{rieke09}. They show even lower far-IR to mid-IR ratios than our low-$z$ ($z$\,$\sim$\,0.8) starburst template, which is based on comparable luminosity sources.  are also closest to the local templates corresponding to about an order of magnitude less luminous local galaxies. They have even lower far-IR to mid-IR ratios than our higher luminosity $z$\,$\sim$\,2 starbursts. 

An important caveat is that while overall the SEDs of our $z$\,$\sim$\,1 and $z$\,$\sim$\,2 starbursts look much like local LIRGs they do show significantly stronger $\sim$\,3-5\um\ continuum emission than seen in the local sources (see Figure\,\ref{fig_compare_templates}). This means that the higher mid-IR continuum is due to relatively more significant AGN contribution than seen in typical LIRGs. Based on our SED fitting the median AGN fraction of the lower-$z$ starbursts is $\sim$\,3\%, and of the higher-$z$ starbursts it is $\sim$\,6\%. This is not much overall, but is significant in the mid-IR regime. 

\begin{figure}[h!]
\epsfig{file=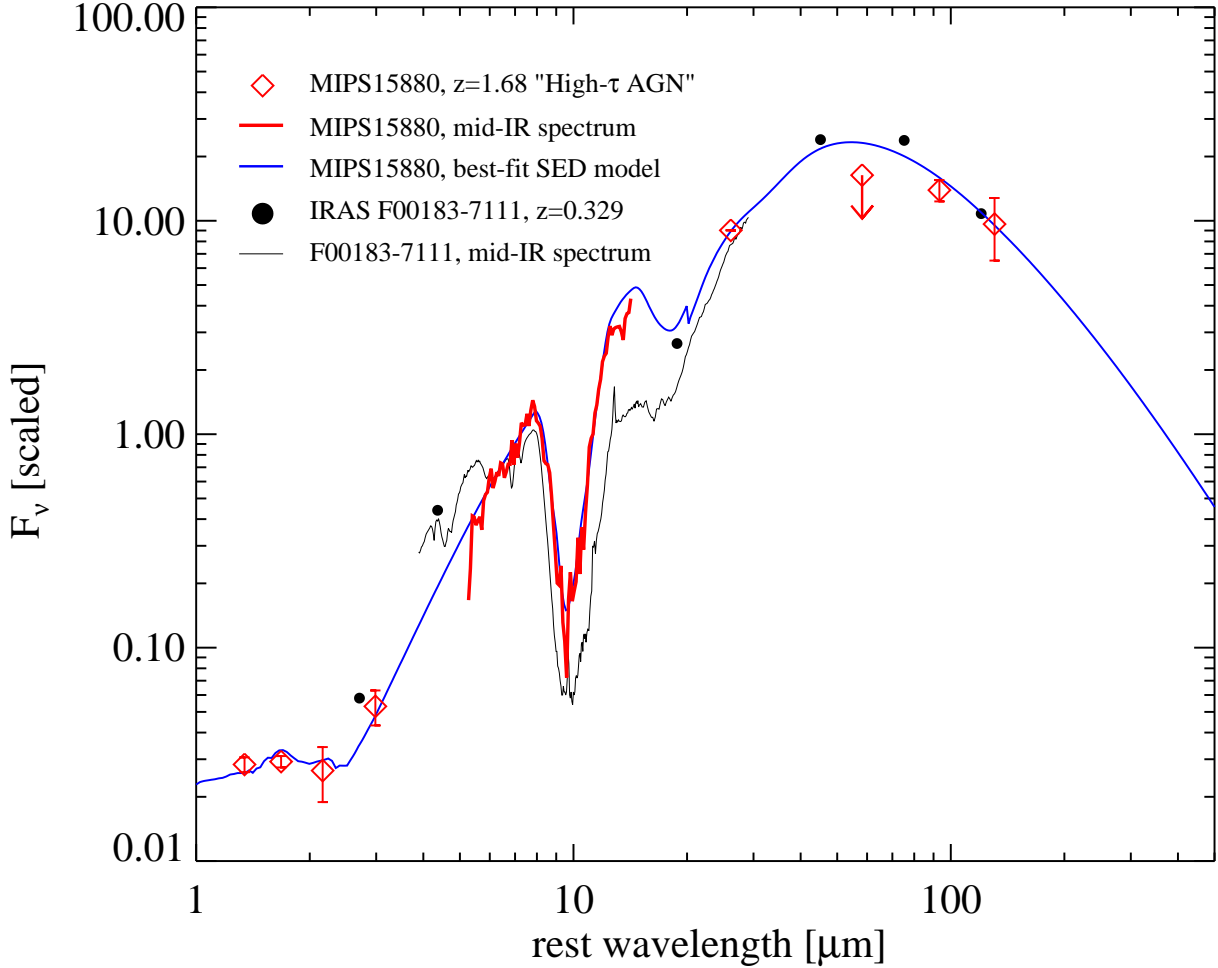,clip=,width=1.0\linewidth}
\caption{A comparison between a particularly strong silicate absorption AGN-dominated source in our sample (MIPS15880, $z$\,=\,1.68) with a close local analog ({\sl IRAS} F00183-7111, $z$\,=\,0.329). Here we show the {\sl Spitzer} IRS mid-IR spectra for both sources \citep{spoon04,yan07} as well as their broadband photometry (where the data for the local source come from NED). The blue solid curve shows the best-fit SED model for MIPS15880. The two are reasonably similar, except our source is redder in the 3-10\um\ regime, brighter at $\sim$\,10\,--20\um, and there is some uncertainly $\sim$\,160\um\ where the 3\,$\sigma$ upper limit of the {\sl Spitzer} 160\um\ data point for MIPS15880 falls short of the best-fit SED model. \label{fig_compare_f00183}}  
\end{figure}

\begin{figure}[h!]
\epsfig{file=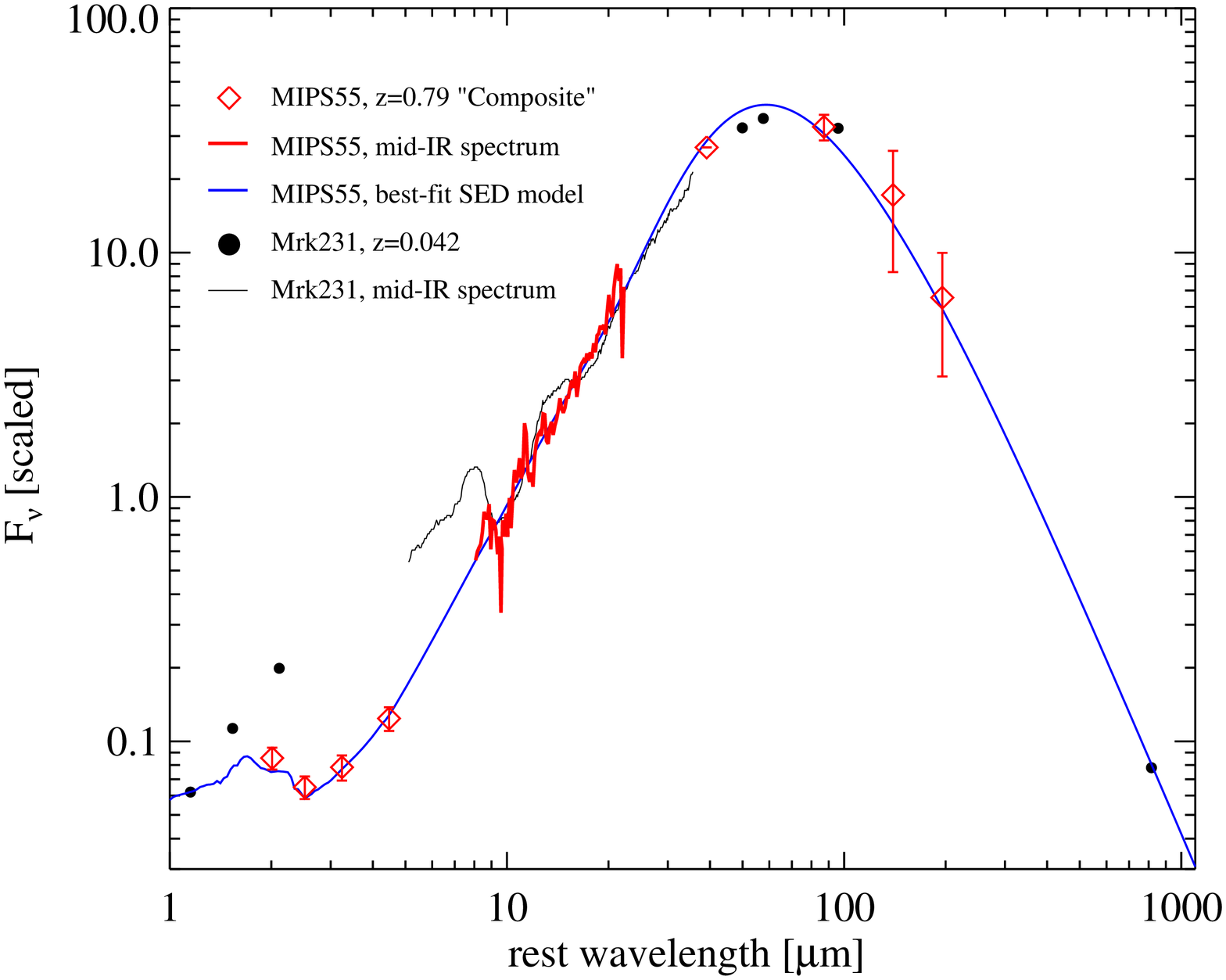,clip=,width=0.5\linewidth}
\epsfig{file=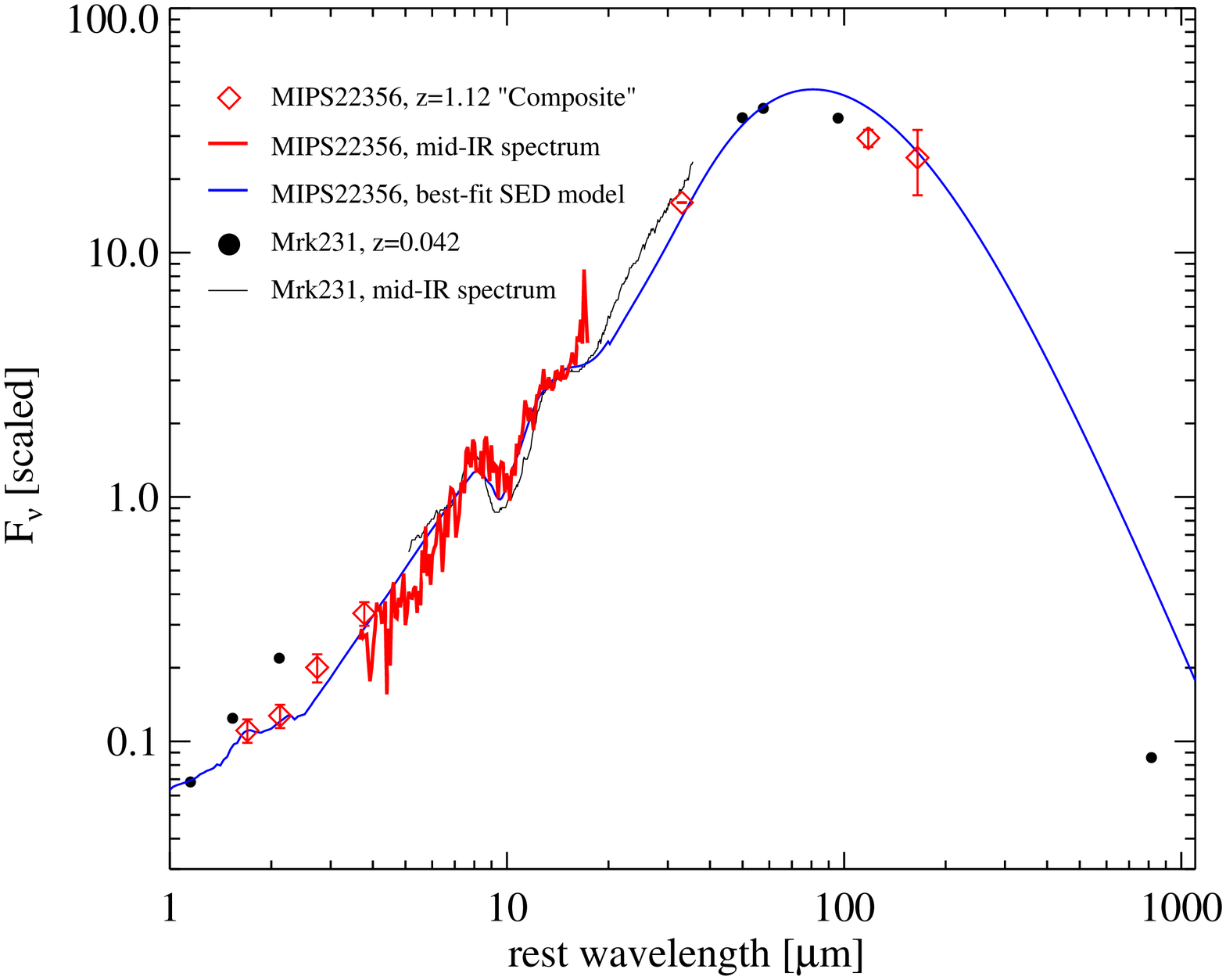,clip=,width=0.5\linewidth}
\caption{Two examples of a clear composite source (MIPS55 and MIPS22356) both of which have no PAH, but a clear far-IR peak. As discussed earlier, these sources are similar to the local warm ULIRG Mrk231. Here we show a more detailed comparison (as in Figure\,\ref{fig_compare_f00183}) showing that indeed these sources' SEDs are indeed very similar to Mrk231, although there are discrepancies especially at $<$\,10\um.  \label{fig_compare_hii}}
\end{figure}

\subsection{Is our interpretation of the nature of the mid-IR AGN correct? \label{sec:miragn}}

When we consider the SED classes shown in Figure\,\ref{fig_compare_templates}, there is little doubt that the strong-PAH, strong far-IR emission sources are starburst-dominated, although likely with non-negligible AGN contribution, or that the low-$\tau$ AGN sources are essentially pure AGN (that compare well with local analogs). However, the nature of the high-$\tau$ AGN sources which are nearly exclusively found at $z$\,$>$\,0.9 in our sample is less unambiguous as is the nature of the "composite" sources. The high-$\tau$ AGN template does not look like any classic AGN population including optical quasars, Seyfert 1s, or local radio galaxies all of which display weak silicate absorption, if any \citep{sturm06,ogle06,hao07}. The existence of such deep silicate absorption feature sources at $z$\,$\sim$\,1\,--\,2 was one of the major discoveries of the {\sl Spitzer} IRS, and although somewhat uncertain still, the general opinion is that these are obscured quasars \citep{houck05,polletta08,sajina09,georgantopoulos11}. While these sources do not look like the {\it typical} local ULIRGs, which sources have predominantly starburst-like mid-IR SEDs (see Section\,\ref{sec:sedevol}), still, the best local analogs to these sources are to be found among the local ULIRGs \citep[see][]{sajina09}. As an example, in Figure\,\ref{fig_compare_f00183}, we look at a source in our sample with particularly strong silicate absorption (MIPS15880) which is reasonably similar in its overall SED to the well studied deep silicate absorption source {\sl IRAS} F00183-7111, which is believed to be largely AGN powered \citep{spoon04}. While extremely rare, such local analogs to our high-$\tau$ sources can indeed be found. MIPS15880 specifically, is also a double-lobed radio galaxy \citep{sajina07b}, which at least supports the presence of an AGN, although not its AGN dominance in the infrared. The overall high levels of obscuration in these sources are consistent with their X-ray non-detections \citep[see e.g.][]{bauer10}. 
 
As for the composite objects, there are two possibilities. The first is that these sources are pure starbursts, but ones with unusually weak PAH emission -- this property is expected of galaxies dominated by their HII regions, such as very young compact starbursts. The dwarf galaxy NGC1377 is the best studied example \citep{roussel06}. We find that both the spectrum of NGC1377 and a theoretical ultra compact HII region template \citep{groves08} are too steep in the mid-IR compared with our sources. Essentially, we can use the 15/5\um\ color as a means of discriminating between HII regions and AGN emission, as first proposed by \citet{laurent00}. Looking back at Figure\,\ref{fig_diag}, it is clear that, the $\log(L_{15}/L_{5.8})$ color is essentially the same for the AGN and composite objects and is only marginally stepper for the starburst sources. 

The other possibility is that, by contrast, these sources are nearly fully powered by AGN, but ones whose emission is significantly cooler (emission region is more extended) than seen in local quasars. Figure\,\ref{fig_compare_hii} addresses the first possibility by comparing our $z$\,$>$\,0.9 composite-source template with both the theoretical ultra compact HII region template of \citet{groves08} and the broadband SED of NGC1377.  The mid-IR SED of our composite objects is significantly shallower than both the UCHII template and NGC1377. Essentially, we are using the 15/5\um\ color as a means of discriminating between HII regions and AGN emission, as first proposed by \citet{laurent00}. Looking back at Figure\,\ref{fig_diag}, it is clear that, the $\log(L_{15}/L_{5.8})$ color is essentially the same for the AGN and composite objects and is only marginally stepper for the starburst sources.  Since it is clear that the classification of the composite sources is the most uncertain (especially for those without good far-IR detections), here we want to examine in highlight the SEDs of two "ideal" examples -- i.e. sources that show no PAH in their mid-IR spectra, but have strong and unambiguous far-IR detections. These examples are shown in Figure\,\ref{fig_compare_hii}. It is clear that, indeed, Mrk231 is a good local analog. Mrk231 itself is known to host an AGN, which dominates in the mid-IR, but is believed to derive $\sim$\,70\% of its overall IR power from star-formation \citep{farrah03}, consistent with our definition of a composite source. 

\subsection{Trends with SED and morphology}

A cross analysis of IR SEDs and morphologies is key to testing our models of galaxy evolution. In \citet{zamojski11}, we address this point; however, using only the mid-IR AGN/SB classification of these sources.  A key finding there concerned our strong-PAH, high-$z$ sources. The implied far-IR luminosities of these sources (accurately measured luminosities were not available until the analysis in this paper) placed them easily in the ULIRG regime, for which we would expect to see the late stage of a major merger, given local analogs.  Instead, these sources showed a mix of morphologies in their rest-frame optical images\footnote{This morphological analysis uses {\sl HST} NICMOS $H$-band data. } , including many disturbed disks, often in the early stages of a merger. This is consistent with recent results suggesting that at moderate to high redshifts, ULIRG-like luminosities can indeed be reached in the early stages of a merger (see Section\,\ref{sec:intro}).  

\begin{figure}[h!]
\centering
\epsfig{file=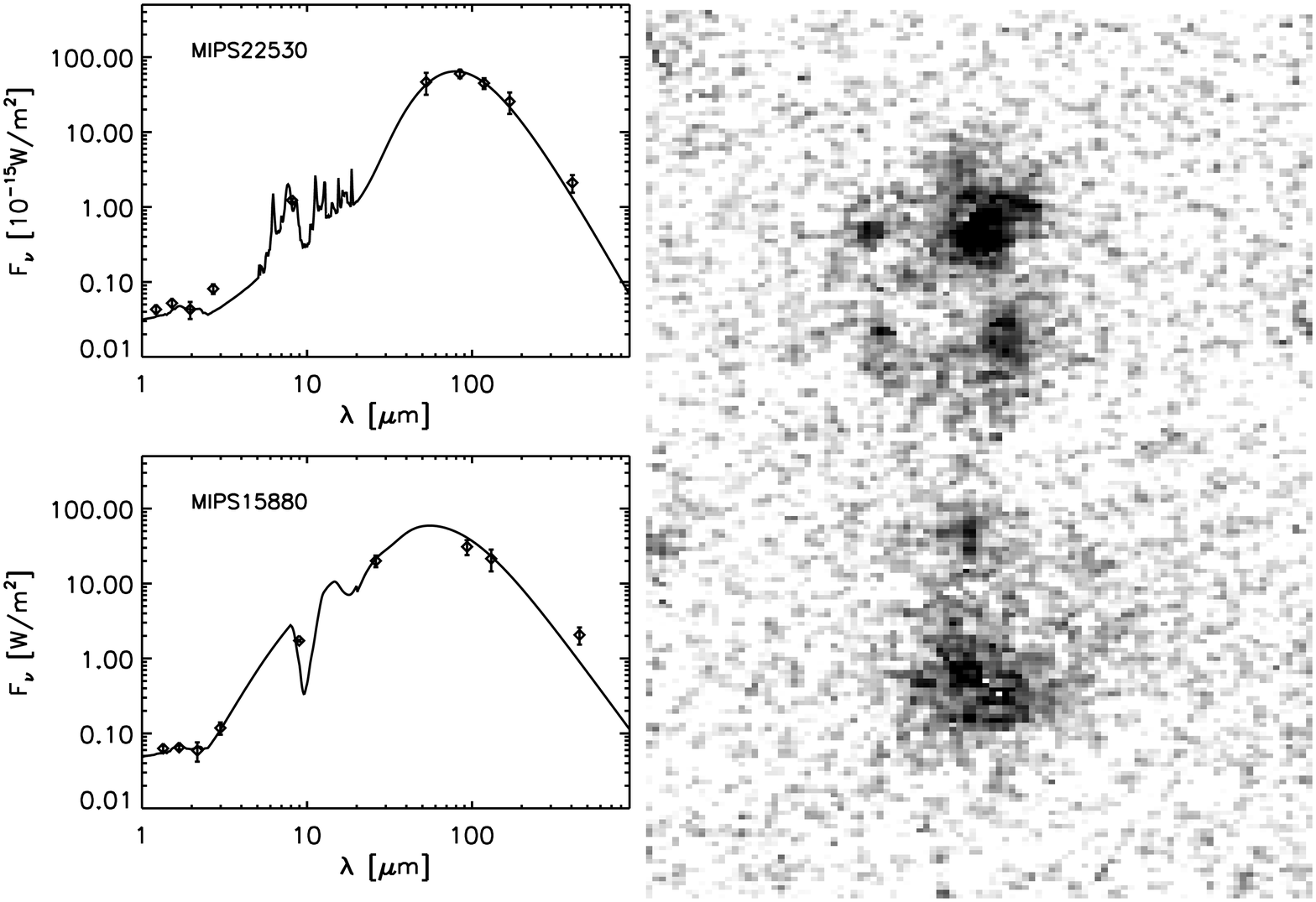,clip=,width=0.7\linewidth}\\
\caption{Examples of a disk-dominated $z$\,$\sim$\,2 starburst (MIPS22530, $z$\,=\,1.96) and a disk-dominated $z$\,$\sim$\,2 AGN (MIPS15880, $z$\,=\,1.68). The left-hand panels show the broad-band photometry for each galaxy along with the best-fit SED.   The right-hand panels show the {\sl HST} NICMOS images (cut to $\sim$\,3$^{\prime\prime}$\,$\times$\,3$^{\prime\prime}$ boxes).  The classification of these galaxies as "disks" is based on surface-brightness profile fitting done by \citet{zamojski11}. \label{fig_morph_examples}}  
\end{figure}

We find that the conclusions of \citet{zamojski11} remains true even when we extend the analysis to the full IR-based AGN/SB classification. Effectively $\sim$\,60\% of the $z$\,$>$\,1.1 starburst-dominated sources show predominantly disk morphologies and are typically in the early stages of a merger (e.g. close pair).  Even more surprisingly, the fraction of disks does not fall with increasing AGN-to-starburst ratio -- indeed the opposite is observed. More than half of the AGN-dominated systems are disks as well.  In Figure\,\ref{fig_morph_examples} we show examples of both a starburst-dominated and an AGN-dominated $z$\,$\sim$\,2 source that are both classified as disks in a "pre-merger" stage in \citet{zamojski11}, evidence for which is seen in the presence of companions and the somewhat asymmetric shapes.  It is clear; however, from Figure\,\ref{fig_morph_examples}, that these conclusions are so far somewhat tentative since our data are not especially deep, and features such as tidal arcs can easily be hidden in the "disk" profiles.  

\subsection{Radio-loud fraction of high-$z$, dust-obscured AGN}
In \citet{sajina07b}, we found that among the GO1 sample, $\sim$\,40\% of the high $\tau_{9.7}$, $z$\,$>$\,1.6 sources have $L_{1.4GHz}$\,$\geq$\,$10^{25}$W/Hz, in other words are radio-loud. We want to use our larger, and color-unbiased supersample to test whether this finding holds for it as well. We start by looking at the total number of radio-loud sources, which is 18. We find that they are at higher redshifts as expected. The lowest redshift radio-loud source is MIPS8253 which is at $z$\,=\,0.953 and is by far the radio-brightest source in our sample with $S_{1.4GHz}$\,$\sim$\,18mJy and $S_{610MHz}$\,$\sim$\,26\,mJy.  The radio-loud sources predominantly have low PAH equivalent width and indeed high $\tau_{9.7}$. Of the 14 $z$\,$>$\,1.6 sources, 10 have $\tau_{9.7}$\,$>$\,1.0. This represents $\sim$\,30\% of all $z$\,$>$\,1.6 $\tau_{9.7}$\,$>$\,1.0 (this fraction is essentially the same regardless of whether or not we include or exclude the few high EW7.7 sources). This suggests that indeed a high fraction of the $z$\,$>$\,1.6 dust-obscured $F_{24}$\,$>$\,0.9\,mJy sources are radio-loud. 

If we interpret these high silicate absorption, AGN-dominated sources as being in a transition state before the "blowout" of their dusty cocoon \citep{hopkins08}, then this supports the view that the development of radio AGN contributes to the feedback processes behind this "blowout". In powerful high-$z$ radio galaxies, it is estimated that the radio jets provide sufficient mechanical energy to drive the observed high speed outflows \citep{nesvadba06}. It is as yet unclear to what degree this may be the case in less extreme sources as the ones discussed here. It does however imply that the development of radio-mode AGN is in some way related to the dusty phase of a quasar's evolution. 

\subsection{Applications of our mid-IR source-based SED templates}

A key outcome of this paper is making public SED templates based on mid-IR selected high-$z$ starbursts and AGN. The first application to these templates is in the interpretation of the 22\um-bright sources detected by {\sl WISE} \citep[{\sl Wide-field Infrared Survey Explorer};][]{wright10}. For example, our high-$\tau_{9.7}$ AGN template has been found to be well matched to the SEDs of higher-$z$ {\sl WISE} sources with available {\sl Herschel} data (Yan et al. 2012, in prep.). Given the flux limits of {\sl WISE}, these sources are at the very tip of the luminosity function $>$\,$10^{13}$\lsun, with our relatively 24\um-bright sources being indeed their closest analogs in the literature. Our templates are also a key step toward a more realistic treatment, in future galaxy population synthesis models, of high redshift galaxy SEDs, especially those dominated by AGN or constituting AGN-starburst composites. In that regard, our study is complementary to a similar recent study (Kirkpatrick et al, 2012, ApJ, in press)  where the properties of fainter 24\um\ sources (typically $F_{24}$\,$\sim$\,0.2\,--\,0.6\,mJy) are examined using {\sl Spitzer} mid-IR spectra and {\sl Herschel} PACS and SPIRE data. In the near future, we intend to combine the results of the two studies in order to produce an SED library that samples the luminosity-redshift space much better than either study by itself. 

\section{Summary \& Conclusions}

In this paper, we combine {\sl Spitzer} and {\sl Herschel} data to study the infrared SEDs  of a sample of 191 $F_{24}$\,$>$\,0.9\,mJy sources in the redshift range 0.3\,--\,2.8 and with derived IR luminosities in the range $10^{11.0}$\,--$10^{13.2}$\lsun. This is the largest uniformly-selected sample of high-$z$ sources with mid-IR IRS spectra, which provide redshifts and spectral classification.  The majority (60\%) of our sources are detected in the nearly confusion limited 250\um\ map of the xFLS obtained as part of the HerMES survey. The 350\um\ and 500\um\ detectability is progressively lower, as expected. These legacy SPIRE data, combined with targeted MIPS70\um\ photometry (69\%\ detected) and some MIPS160 and MAMBO1.2mm data allow us to accurately determine the total IR power output of our sources. Combining these data with {\sl Spitzer} IRS spectra and IRAC photometry allow us  to fit composite empirical models  from which the relative contribution of AGN and star-formation to the total power output can be determined. Our key conclusions are:

\begin{enumerate}
\item  This full IR SED analysis confirms earlier results that this 24\um-selected sample consists of a heterogeneous mix of starburst-dominated sources (30\%, predominantly at lower redshifts), AGN-dominated sources (23\%) as well as composites, including starbursts with AGN-like mid-IR spectra (47\%). 
\item Comparing our derived AGN fractions with various mid-IR spectral diagnostics, we find that the 7.7\um\ PAH equivalent width and the 30-to-15\um\ color are the best predictors of the relative AGN strength in a given galaxy. 
\item The silicate absorption feature alone is not a good predictor of the AGN fraction because many strong-PAH sources are accompanied by strong silicate absorption as well. However, among our $z$\,$>$\,1.2 AGN-dominated sources nearly 2/3 show strong silicate absorption. These sources are also more likely to be radio-loud compared to low-$\tau_{9.7}$ AGN-dominated sources.  
\item The mid-IR SEDs of our starburst sources tend to be more like those of local LIRGs than local ULIRGs. More specifically the local ULIRGs have redder 30-to-15\um\ colors and deeper silicate absorption features than seen in either our sample or local LIRGs. This is also consistent with our earlier morphological analysis \citep{zamojski11}, suggesting that our strong-PAH sources (even those at $z$\,$\sim$\,2 and with $L_{\rm{IR}}$\,$\sim$\,$10^{13}$\lsun) tend to be in an earlier merger stage than typical of local ULIRGs. This supports earlier results based on longer wavelength selected samples \citep{huynh10,seymour10,muzzin10}. 
\item We make public SED templates derived from our $z$\,$\sim$\,0.3\,--\,3.0 mid-IR bright sources which are representative of such high redshift starbursts,  obscured AGN, and starburst-AGN composites. These are already being used in the interpretation of the high-$z$ sources detected in the all sky mid-IR {\sl WISE} survey. We hope for our templates to help improve future infrared galaxy evolution models.
\end{enumerate}

\acknowledgements
Most of all, we are grateful to the HerMES team (PI Seb Oliver) for the excellent public dataset which we use extensively here. We are also grateful to the anonymous referee for their careful reading and helpful suggestions, which have greatly improved the content and presentation of this paper. This  paper has benefited from very helpful discussions on IR SEDs and their interpretation with Brent Groves, Patrik Jonsson, and Chris Hayward. We are very grateful to Sylvain Veilleux for providing us with the IRS spectra of local ULIRGs and PG quasars; to Andreea Petric for providing us with the average IRS spectra of GOALS LIRGs; and to Brent Groves for providing us with the Ultra Compact HII region SED template.  We make use of the public clumpy torus models of Maia Nenkova. Overall, we use the wide range of archival data available in the xFLS including redshifts and photometry and are grateful to all the people who have made these data available. 
This work is based in part on observations made with the {\sl Spitzer} Space Telescope, which is operated by the Jet Propulsion Laboratory, California Institute of Technology under a contract with NASA. This paper also makes use of {\sl Herschel} data. {\sl Herschel} is an ESA space observatory with science instruments provided by European-led Principal Investigator consortia and with important participation from NASA. 
Support for this work was provided by NASA through an award issued by JPL/Caltech. 

\bibliography{seds.bib}

\begin{center}

\begin{deluxetable}{lcccc}
\tablecolumns{5}
\tablewidth{4in}
\tabletypesize{\tiny}
\tablecaption{\label{table_prelims} Sources Names, Redshifts, Mid-IR fitting results and references} 
\tablehead{\colhead{Source} &  \colhead{$z$} & \colhead{EW7.7} & \colhead{$\tau_{9.7}\tablenotemark{\dagger}$} & \colhead{Notes/References\tablenotemark{\star}}}
 \startdata
MIPS34 & 0.646?\tablenotemark{a} &              $<$0.15 &   0.0 &                     AGN,L07, D09 $z_{\rm{MIR}}$=1.38 \\
MIPS39 &   2.55 &              $<$0.15 &   2.2 & AGN,D09 $z_{\rm{MIR}}$=2.42, L07 Ly$\alpha$?-based z \\
MIPS42 &   1.95 &              $<$0.15 &   0.8 &                                                      \\
MIPS45 &   0.80 &              $<$0.15 &   0.2 &                                          AGN,L07,H09 \\
MIPS55 &  0.791 &        0.67$\pm$0.05 &   0.1 &                                           galaxy,P06 \\
MIPS78 &   2.65 &              $<$0.15 &   1.8 &                                                      \\
MIPS110 &  1.000 &              $<$0.24 &   0.5 &                                             AGN, S08 \\
MIPS133 &   0.90 &              $<$0.15 &   1.4 &                                                      \\
\enddata
\tablenotetext{\dagger}{This is the $\tau_{9.7}$ that results from the fitting as described in \citet{sajina07a}. To convert to the observed depth of the silicate absorption feature, $\tau_{sil}$, these values need to be divided by 1.4.}
\tablenotetext{\star}{Here the references given are the origin of the optical/near-IR spectroscopic redshifts where C06 is \citet{choi06}; L07 is \citet{lacy07}; L11 is \citet{lacy11}; D09 is \citet{dasyra09}; P06 is \citet{papovich06}; MS06 is \citet{ams06}, S08 is \citet{sajina08}.}
\end{deluxetable}
\end{center}

\begin{center}
\begin{deluxetable}{ccccc ccccc ccccc}
\rotate
\tablecolumns{15}
\tablewidth{8.3in}
\tabletypesize{\tiny}
\tablecaption{\label{table_fluxes} Broadband photometry\tablenotemark{a}} 

\tablehead{\colhead{Source} &  \colhead{$F_{0.64}$} & \colhead{$F_{3.6}$} &   \colhead{$F_{4.5}$} & \colhead{$F_{5.8}$} & \colhead{$F_{8.0}$} & \colhead{$F_{24}$} & \colhead{$F_{70}$} &  \colhead{$F_{160}$} & \colhead{$F_{250}$} & \colhead{$F_{350}$} & \colhead{$F_{500}$} &  \colhead{$F_{1200}$} & \colhead{$F_{20\rm{cm}}$} & \colhead{$F_{610\rm{MHz}}$} \\
\colhead{} &  \colhead{$\mu$Jy} & \colhead{$\mu$Jy} &   \colhead{$\mu$Jy} & \colhead{$\mu$Jy} & \colhead{$\mu$Jy} & \colhead{mJy} & \colhead{mJy} & \colhead{mJy} & \colhead{mJy} & \colhead{mJy} & \colhead{mJy} & \colhead{mJy} & \colhead{mJy} & \colhead{mJy}  }
 \startdata
MIPS34 &  11.7 & 446$\pm$45 & 680$\pm$68 &  980$\pm$101 & 1414$\pm$143 &  5.42$\pm$0.06 &   14.4$\pm$2.6 & $<$48 & 12$\pm$5 & $<$16 & $<$18 & --- & --- & --- \\
MIPS39 &  1.52 &  58$\pm$7 & 135$\pm$15 &  323$\pm$37 &  972$\pm$101 &  5.10$\pm$0.07 &   10.2$\pm$1.7 & $<$60 & 15$\pm$6 & 14$\pm$5 & $<$18 & -0.14$\pm$0.65 & --- & --- \\
MIPS42 & $<$\,0.190 & $<$9 &  30$\pm$2 &  103$\pm$16 &  680$\pm$15 &  4.96$\pm$0.06 &   13.2$\pm$3.4 & $<$41 & $<$20 & $<$19 & $<$24 & 0.43$\pm$0.62 & --- & --- \\
MIPS45 &  12.0 & 523$\pm$53 & 819$\pm$83 & 1130$\pm$117 & 1800$\pm$182 &  4.84$\pm$0.07 &   18.5$\pm$3.8 & $<$42 & $<$21 & $<$21 & $<$21 & 0.16$\pm$0.65 & 0.57$\pm$0.03 &  1.3$\pm$0.18 \\
MIPS55 &  15.5 & 190$\pm$19 & 144$\pm$15 &  174$\pm$21 &  276$\pm$31 &  4.13$\pm$0.06 &   59.9$\pm$8.7 &  73$\pm$20 & 38$\pm$7 & 14$\pm$6 & $<$23 & --- & 0.34$\pm$0.03 & --- \\
MIPS78 & $<$\,0.190 & $<$12 &  39$\pm$5 & $<$72 &  268$\pm$19 &  3.20$\pm$0.06 & $<$3.9 & $<$51 & $<$20 & $<$19 & $<$21 & -0.25$\pm$0.62 & --- & --- \\
MIPS110 &  1.98 &  44$\pm$4 &  44$\pm$2 &   72$\pm$19 &  200$\pm$15 &  1.81$\pm$0.06 &   12.1$\pm$3.3 & $<$36 & $<$19 & $<$17 & $<$18 & -0.42$\pm$0.55 & 0.37$\pm$0.03 & 0.81$\pm$0.16 \\
MIPS133 &  4.37 &  58$\pm$7 &  54$\pm$4 & $<$66 &  155$\pm$17 &  2.15$\pm$0.06 &   19.9$\pm$3.1 & $<$31 & $<$20 & $<$18 & $<$21 & 0.01$\pm$0.57 & 0.26$\pm$0.01 & 0.73$\pm$0.09 \\
\enddata
\medskip
\tablenotetext{a}{Upper limits are 3\,$\sigma$. }
\end{deluxetable}
\end{center}

\begin{center}
\begin{deluxetable}{c cccc ccccc cccc}
\rotate
\tablecolumns{11}
\tablewidth{8in}
\tabletypesize{\tiny}
\tablecaption{\label{table_lums} SED-fitting Derived Quantities} 
 \tablehead{\colhead{Source} &  \colhead{$\alpha_{\rm{hot}}$} & \colhead{Td\tablenotemark{a}} & \colhead{$\beta$\tablenotemark{a}} & \colhead{$L_{1.6\mu m,stars}$} & \colhead{$L_{5.8\mu m}$} & \colhead{$L_{15\mu\rm{m}}$} & \colhead{$L_{30\mu\rm{m}}$} & \colhead{$L_{1.4GHz}$} &\colhead{$L_{\rm{AGN}}$} & \colhead{$L_{3-1000}$}  & \colhead{SFR} & \colhead{($L_{\rm{AGN}}/L_{\rm{IR}}$)$_{ML}$} & \colhead{Classification\tablenotemark{b}} \\
\colhead{} & \colhead{} & \colhead{K} & \colhead{} & \colhead{\lsun} & \colhead{\lsun} & \colhead{\lsun}  & \colhead{\lsun} & \colhead{$W/Hz$}  & \colhead{\lsun} &  \colhead{\lsun} & \colhead{\msun/yr} & \colhead{\%} & \colhead{} }
 
 \startdata
   MIPS34 &       1.1 &    (50.0) &     (1.5) &     --- &   11.43 &   11.53 &   11.57 & $<$23.17 &   11.79 & 11.99$\pm$0.03 & 62.28 &  72\,$\pm$\,10 &  low-$\tau_{9.7}$ AGN \\
   MIPS39 &       0.8 &      38.6 &     (1.5) &   11.91 &   12.97 &   12.82 &   12.64 & $<$24.58 &   13.28 & 13.28$\pm$0.04 &   --- &  99\,$\pm$\,11 & high-$\tau_{9.7}$ AGN \\
   MIPS42 &       1.4 &    (50.0) &     (1.5) &     --- &   12.61 &   12.73 &   12.70 & $<$24.31 &   13.00 & 13.12$\pm$0.04 & 549.6 &   81\,$\pm$\,7 &  low-$\tau_{9.7}$ AGN \\
   MIPS45 &       0.8 &    (50.0) &     (1.5) &     --- &   11.72 &   11.69 &   11.70 &   24.25$\pm$  0.02 &   12.03 & 12.18$\pm$0.02 & 76.34 &   77\,$\pm$\,9 &  low-$\tau_{9.7}$ AGN \\
   MIPS55 &       2.3 &      62.9 &       1.5 &   11.13 &   11.07 &   11.69 &   12.18 &   23.95$\pm$  0.04 &   12.05 & 12.51$\pm$ 0.1 & 365.0 &   0\,$\pm$\,10 &            Composite? \\
   MIPS78 &       1.1 &    (50.0) &     (1.5) &     --- &   12.75 &   12.75 &   12.65 & $<$24.62 &   13.18 & 13.19$\pm$0.04 & 60.88 &   99\,$\pm$\,3 & high-$\tau_{9.7}$ AGN \\
  MIPS110 &       1.1 &    (50.0) &     (1.5) &   10.73 &   11.31 &   11.73 &   12.04 &   24.28$\pm$  0.04 &   11.39 & 12.27$\pm$0.02 & 279.1 &   14\,$\pm$\,5 &           Starburst-A \\
  MIPS133 &       1.4 &    (50.0) &     (1.5) &   10.77 &   11.02 &   11.56 &   11.88 &   24.10$\pm$  0.03 &   11.50 & 12.23$\pm$0.02 & 238.6 &   15\,$\pm$\,6 &           Starburst-A \\
\enddata
\medskip
\tablenotetext{a}{Parameters in parenthesis are held fixed.}
\tablenotetext{a}{If the best $\chi^2$ AGN fraction and the maximum likelihood AGN fraction differ by $>$\,20\%, which would result in a different classification, then the classification is highly questionable and is marked by "?".}
\end{deluxetable}
\end{center}

\appendix

\section{Source ID confusion \label{sec_match}}
\subsection{In IRAC images \label{sec_irac_blend}}
On average $\sim$\,13\% of our sources have multiple IRAC sources within the MIPS24\um\ beam -- some examples are shown in Figure\,\ref{fig_blended}. \citet{dasyra09} discuss the IRAC source identification for the GO2 sample, and conclude that multiple IRAC ID's (including star contaminants) are found in $\sim$\,15\% of the GO2 sources. For the GO1 sample, there were 6 such sources ($\sim$\,13\%) \citep{sajina07a}, of which 4 sources (MIPS42, MIPS110, MIPS279, MIPS289, and MIPS22661) are in the supersample. For the additional 17 sources, 14 had unambiguous IRAC detections in the xFLS IRAC catalog and we adopt their catalog fluxes. Two of the sources, 12509696 and 19454720 do show faint sources in the IRAC 3.6\um\ image, but not strong enough to be in the 5\,$\sigma$ catalog. We estimate their fluxes separately using aperture photometry as in \citet{sajina07a}. Source 12509696 is found in-between a pair of nearly blended IRAC sources (one strong and one faint). Because of the source's high-$z$ and for consistency with the IRS flux, the fainter IRAC source is the more likely counterpart to the MIPS source.

\subsection{In MIPS and SPIRE images \label{sec_fir_blend}}
The large beams of MIPS70, 160\um\ and SPIRE 250, 350 and 500\um\ lead to confusion due to multiple 24\um\ sources within the beam in $\sim$\,10\,--\,30\% of cases (see Table\,\ref{table_det} for details). The worst here are MIPS160\um\ and SPIRE500\um; however, those are also the bands where the fraction of source detections is lowest.  Figure\,\ref{fig_blended} shows the bulk of the sources suffering from confusion either in IRAC (see above) or more commonly in the far-IR bands. Cases where the MIPS24\um\ sources are sufficiently spaced out so that the SPIRE image appears resolved (e.g. ) are easily dealt with using our custom-written deblending code (see Section\,\ref{sec_debl_code}). APEX, which we use for the MIPS70\um\ and MIPS160\um\ photometry uses PRF fitting and hence does a similar type of deblending to our SPIRE photometry code. We find that sources that appear resolved at 70\um\ have flux densities consistent with the ones we derive from the SPIRE deblending code (the MIPS70 and SPIRE250\um\ beams are comparable). Some more complicated sources (also shown in Figure\,\ref{fig_blended} have to be dealt with on a case by case basis as described below.

\begin{figure}[!h]
\plotone{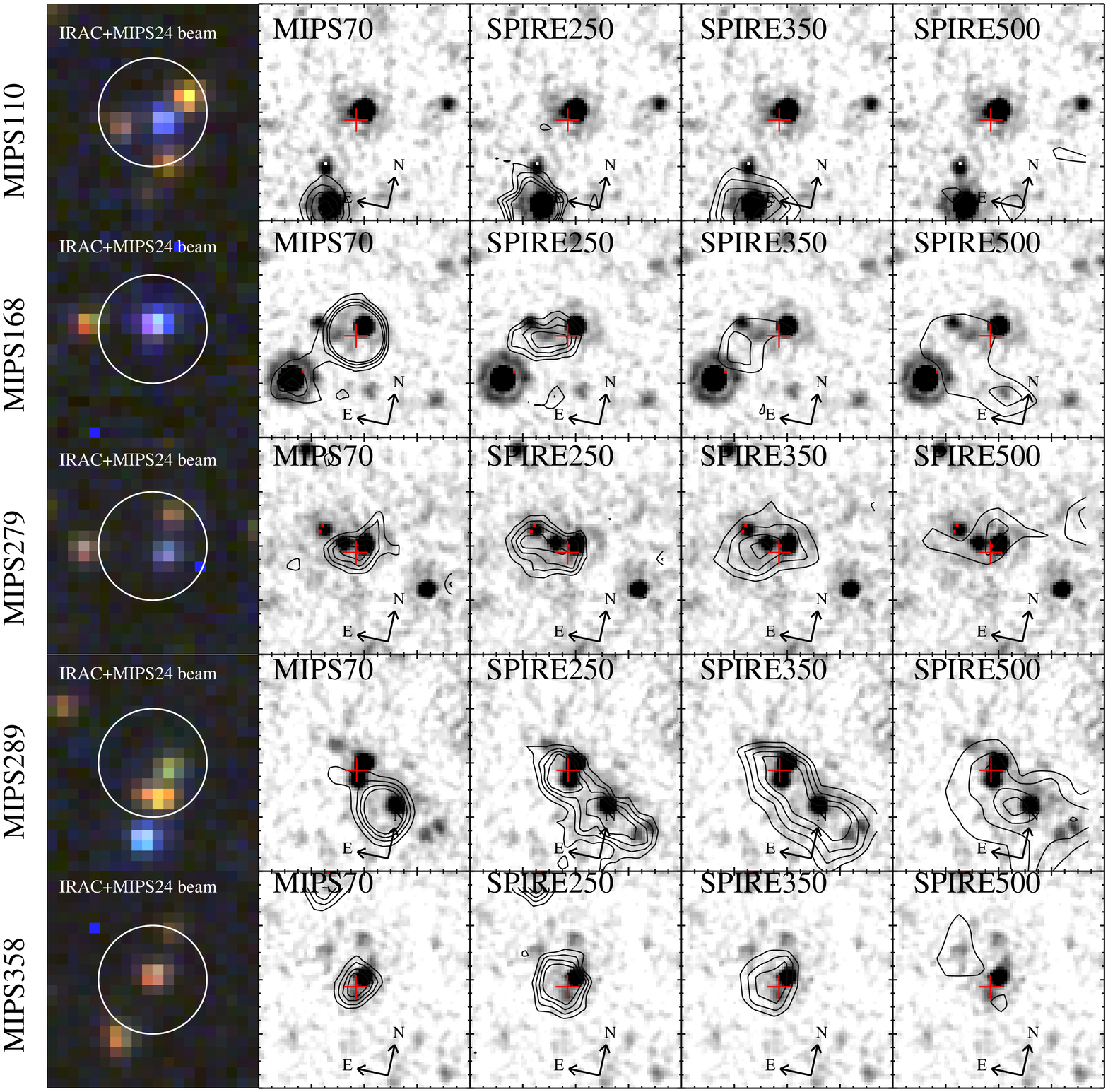}
\caption{Sources with confusion issues either at IRAC, MIPS, or SPIRE. The first panel shows 24"$\times$24" IRAC color cutouts (blue=3.6um,green=4.5um,red=8.0um) with the MIPS24\um\ beam overlaid as a white circle. The rest of the panels show 90"$\times$90" 24\um\ image cutouts overlaid with MIPS70\um\ or SPIRE contours as indicated. The contours are 2,3,4,5, and 6\,$\times$ the confusion level in each band. \label{fig_blended}}
\end{figure}

\setcounter{figure}{11}

\begin{figure}[!h]
\plotone{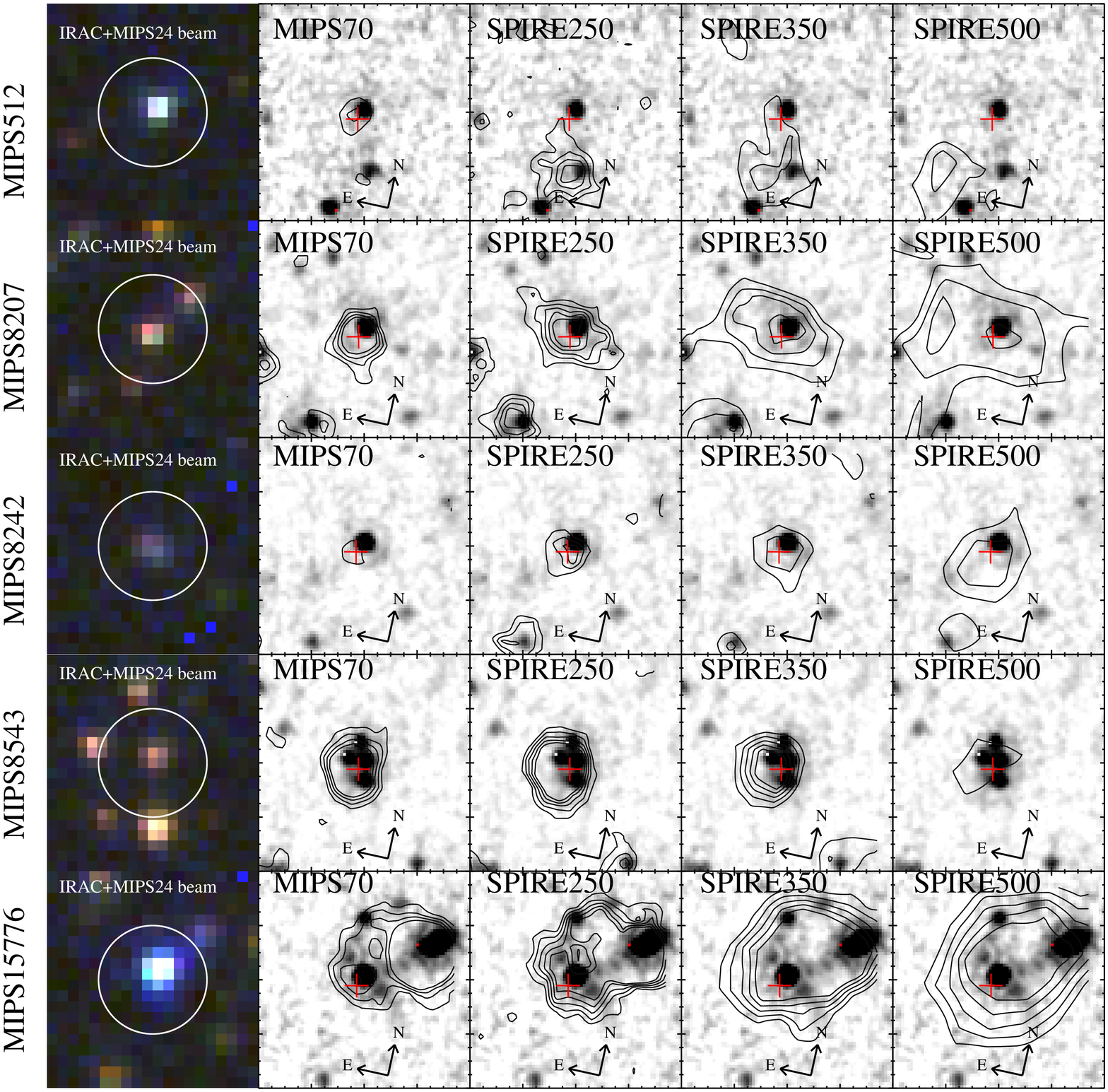}
\caption{{\it Continued}}
\end{figure}

\setcounter{figure}{11}

\begin{figure}[!h]
\plotone{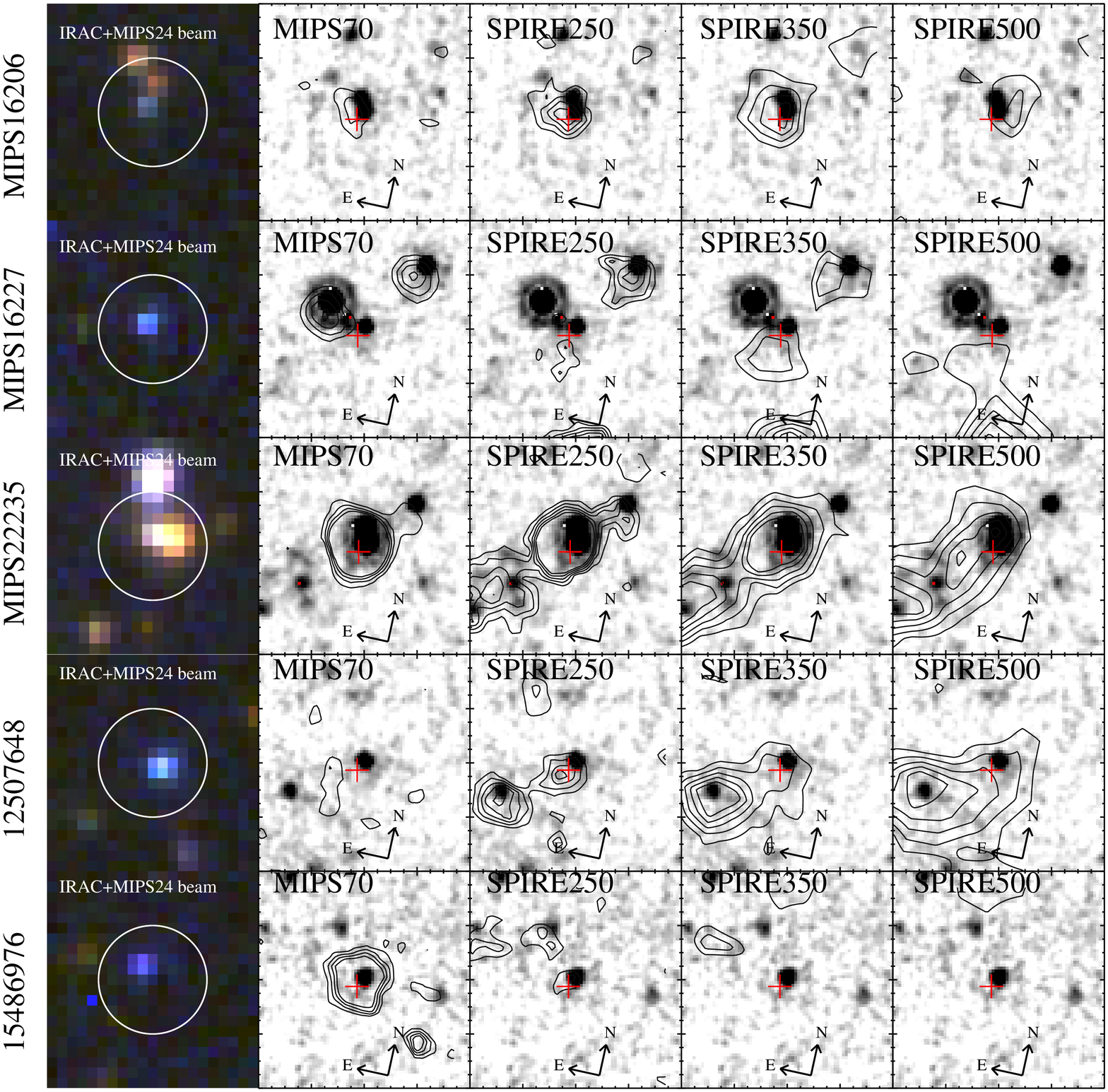}
\caption{{\it Continued}}
\end{figure}

{\bf MIPS168;} It is obvious from Figure\,\ref{fig_blended}, that our target dominates the 70\um.  At 250\um, our source is blended with its neighbor, although it is clearly dominant. The SPIRE deblending code confirms this. The source is detected at 160\um, and from the above, we assume our source dominates the emission.  \\
{\bf MIPS279:} There are three mips24um sources in a row about 10" apart. Following the discussion in \citet{sajina07b} we ascribe our source the full 70\um\ flux. The SPIRE deblending code works reasonably well to separate their contributions to the SPIRE fluxes. See Figure\,\ref{fig_deblend}.\\
{\bf MIPS289:} There are 3 additional MIPS24\um\ sources nearby, including one right next to our source. This is an example, where while the SPIRE deblending code ascribes the bulk of the far-IR emission to our source, over its near neighbor, it is not automatically obvious that this is correct. With two 24\um\ sources this close together, the deblending is less reliable. In this case, we accept the solution as our source has strong PAH emission, and a MAMBO 1.2\,mm detection, and hence its also being a strong far-IR source is reasonable.  \\
{\bf MIPS358:} This source has 24\um\ neighbors at separations of 7.1\arcsec, 7.5\arcsec, 15\arcsec\ and 19\arcsec. However, the emission in all far-IR bands is centered in our source (which is also by far the strongest in MIPS24\um, and is a strong-PAH source). The SPIRE deblending code confirms this by ascribing all the emission to our source. \\
{\bf MIPS512:} This is a source where the SPIRE deblending code suggests a marginal detection at 250 and 350\um; however, the resulting SED seemed unrealistic. Visual inspection of the image suggested that these SPIRE values should be treated as upper limits instead. \\
{\bf MIPS8207:} This is a code where the photometry and resulting SED are reasonable except at 500\um. We treat our derived SPIRE500\um\ flux density as an upper limit due to the uncertain level of confusion within it. \\
{\bf MIPS8242:} This source is detected in SPIRE and MAMBO. The photometry is clean in all bands excepts 500 where the measured flux seems too high given the SED. While there is not clear single culprit for this excess, an examination of the 24\um\ image suggests this area is particularly rich in faint 24\um\ sources, therefore, the effective confusion noise is higher than usual. We estimate an additional confusion-driven rms from a small box right next to the source of 8.9mJy which added to the cleaner box rms of 7.8 means a total rms of 11.5mJy. \\
{\bf MIPS8543:} There are four 24\um\ sources (with 0.56, 0.64, 0.80, and 0.94mJy) found in close proximity to our source -- visual inspection and given their very similar IRAC colors suggests that this may even be a small group. SPIRE deblending suggests that our source contributes about half of the total 250\um\ emission (and similar for the other SPIRE bands). We therefore split the APEX derived MIPS70 and MIPS160\um\ fluxes by half. This source also has a $S_{850}$\,=\,7.0\,$\pm$\,2.3\,mJy sub-mm counterpart \citep{frayer04}. The SCUBA\footnote{Sub-mm Common User Bolometer Array \citep{holland98}.} 15\arcsec\ beam is centered between our source and the $S_{24}$\,=\,0.56\,mJy source.  Therefore we split the SCUBA flux by half as well. Overall, this results in a reasonable SED with detections in all bands; however, the far-IR photometry for this source should be treated with caution. \\
{\bf MIPS15776:} This source is in an area with several other MIPS24\um\ sources and strong far-IR emission.  Here the combination of APEX photometry for MIPS70 and the SPIRE deblending code gives believable 70\um\ and 250\um\ flux densities. However the 350 and 500\um\ fluxes are unrealistically high. Here we adopt the derived 350 and 500\um\ flux density values as 3\,$\sigma$ upper limits. \\
{\bf MIPS16206:} There is another 24\um\ source of comparable flux 5.6\arcsec\ away from our source. Here the 70\um\ is actually centered on the other source, but has an elongated structure suggesting that our source contributes as well. The flux density for the 70\um\ source is adopted as an upper limit for MIPS16206. \\
{\bf MIPS16227:} We looked at this source in particular because it showed an anomalously high 350\um\ (this is seen both by running the SPIRE deblending code and just by reading off the pixel values from the SPIRE images). Here, while we formally have a 3\,$\sigma$ detection at 350\um\ -- examination of the images by eye suggests that while the source of this emission is not obvious it is unlikely to be our source. We therefore treat this 350\um\ flux as an upper limit. \\
{\bf MIPS22235:} This source is partially blended even in the 24\um\ image (resolved into three sources with 3.16, 1.62, and 0.4\,mJy in the 24\um\ catalog). We find that we need to sum up the 24\um\ flux of the whole system to match the flux levels of the IRS spectrum.  The spectrum does not suggest multiple redshifts (it agrees with the optical spectroscopic $z$\,=\,0.4).  The APEX derived 70\um, and the SPIRE deblending code based photometry lead to a reasonable looking SED with $\sim$\,40\,K dust temperature. Here, similar to MIPS289 we adopt the fluxes as derived in the standard procedure; however, caution that there's a possibility of additional source(s) contributing to this far-IR emission. \\
{\bf 12507648:} This source appears that it would be easily deblended; however, the deblending procedure resulted in too high values for SPIRE350 and SPIRE500\um\ -- effectively resulting in a flat far-IR spectrum. A possible explanation here is that due to its being next to a very bright source, we are affected by its first Airy ring. Our code uses Gaussian profiles rather than the proper PSF's, hence it does not cope well with this situation. Here we decide to read off the pixel values (in Jy/beam) directly from the image. This resulted in a more reasonable looking SED. \\
{\bf 15486976} This source has unusually high 70um flux but is not detected at 160\um\ or in any of the SPIRE bands. Visual inspection of the image shows that it is in a particularly noisy area of the 70um image. It may also be the case that the 70\um\ emission is affected by poorly cleaned artifacts. We treat this 70\um\ point as an upper limit. 

\subsection{SPIRE deblending procedure \label{sec_debl_code}}
\begin{figure}[!h]
\plotone{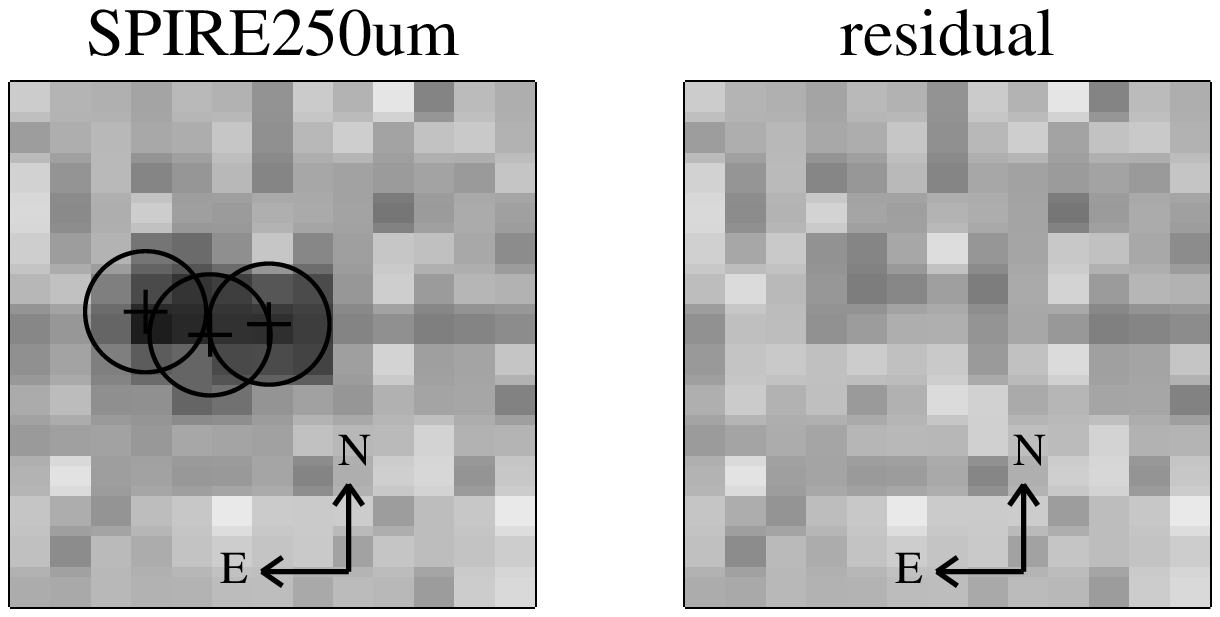}
\caption{An example of source deblending where the image shown is the SPIRE 250um image centered on MIPS279. The crosses mark the positions of 24um sources and the circles are the 18.1\asec\ FWHM of the SPIRE 250um beam. The image on the right is the residual.  \label{fig_deblend}}
\end{figure}

Our SPIRE photometry and source deblending is performed with a custom written code which uses the 24um source positions and 2D Gaussian profiles with the corresponding SPIRE FWHM values. For closely spaced sources, we fix the positions to the 24um source positions; however, for isolated sources, we center on the 250\um\ image (using the {\sc idl} procedure {\sc gcntr}) before proceeding with the Gaussian profile fit. We also weight with the image error arrays in order to avoid being biased by bad pixels or noisy patches on the sky. We find that the above procedure gives good results both for isolated sources as well as blended sources that are sufficiently separated for the SPIRE source to appear elongated (see examples in Figure\,\ref{fig_blended}. As an example, in  Figure\,\ref{fig_deblend} we show the results of the deblending of MIPS279 from its two nearby neighbors. However, multiple 24\um\ sources well within a SPIRE beam are not reliably separated by this procedure. This is especially true at 500\um\ where the FWHM is 36.3\asec. This can result in automatic 500\um\ fluxes that are well in excess of what extrapolation from the rest of the SED would suggest. We examine by eye all SEDs and determine 10 cases with unrealistic 500\um\ fluxes ($\sim$\,1/3 of all sources with detections in this band). 

\section{MCMC fitting and associated uncertainties\label{sec_mcmc}}

Here we summarize the key points of our Markov Chain Monte Carlo (MCMC) fitting approach which is described in more detail in \citet{sajina06}, and fundamentally follows the Metropolis-Hastings algorithm \citep{metrop53,hastings70}. The basic idea behind MCMC is to effectively sample the joint posterior probability distribution for all model parameters by building up chains of random guesses of parameter values, where each successive guess is chosen from some narrow proposal distribution (in our case a multivariate Gaussian), around the previous chain link. This proposed move to a new set of parameters is accepted or rejected according to some criterion, which both pushes the chain toward higher probability regions, and allows for some random deviation from the straight gradient descent-type path. Defining $\Delta \chi^2$ as the $\chi^2$ difference between the current trial step and the previous accepted step, we accept a move if: 1) $\Delta \chi^2$\,$<$\,0 or a uniform random number, $u$ between 0 and 1 meets the criterion $u<e^{-\Delta \chi^2}$. This procedure both finds the best-fit set of parameters, but also keeps chains of "guesses" that effectively sample the posterior probability distribution for each parameter. 

The key best-fit (lowest $\chi^2$) SED parameters are given in Table\,\ref{table_lums}. However, given the wide range of far-IR coverage (including lack thereof) as well as parameter degeneracies such as optically-thin, lower temperature solutions can look much like optically-thick higher temperature solutions as well as the well known $T-\beta$ degeneracy \citep[e.g.][]{sajina06}, we caution against over-interpreting especially our dust temperature values. On the other hand, our IR luminosity measurements are much less sensitive to such degeneracies being ultimately simply a function of the mid-IR plus far-IR continuum emission. To understand the uncertainties therein, we construct the posterior probability distribution of $L_{\rm{IR}}$ as the histograms of all chain values where $\chi^2<\chi^2_{min}+10$, which avoids the initial burn-in period\footnote{Since successive links on the chain are highly correlated, chains are always thinned before the posterior probability distribution is constructed -- here we adopt a factor of 30 thinning.}.  Essentially, the peak of this histogram represents the maximum likelihood $L_{\rm{IR}}$ for a given source and the width represents its uncertainty (computed as the range that encompasses 68\% of the points). These maximum likelihood estimates are nearly always essentially the same as the best $\chi^2$ estimates (which we adopt), and in the few exceptional cases, are within the errors quoted.  

\begin{figure}[h!]
\plottwo{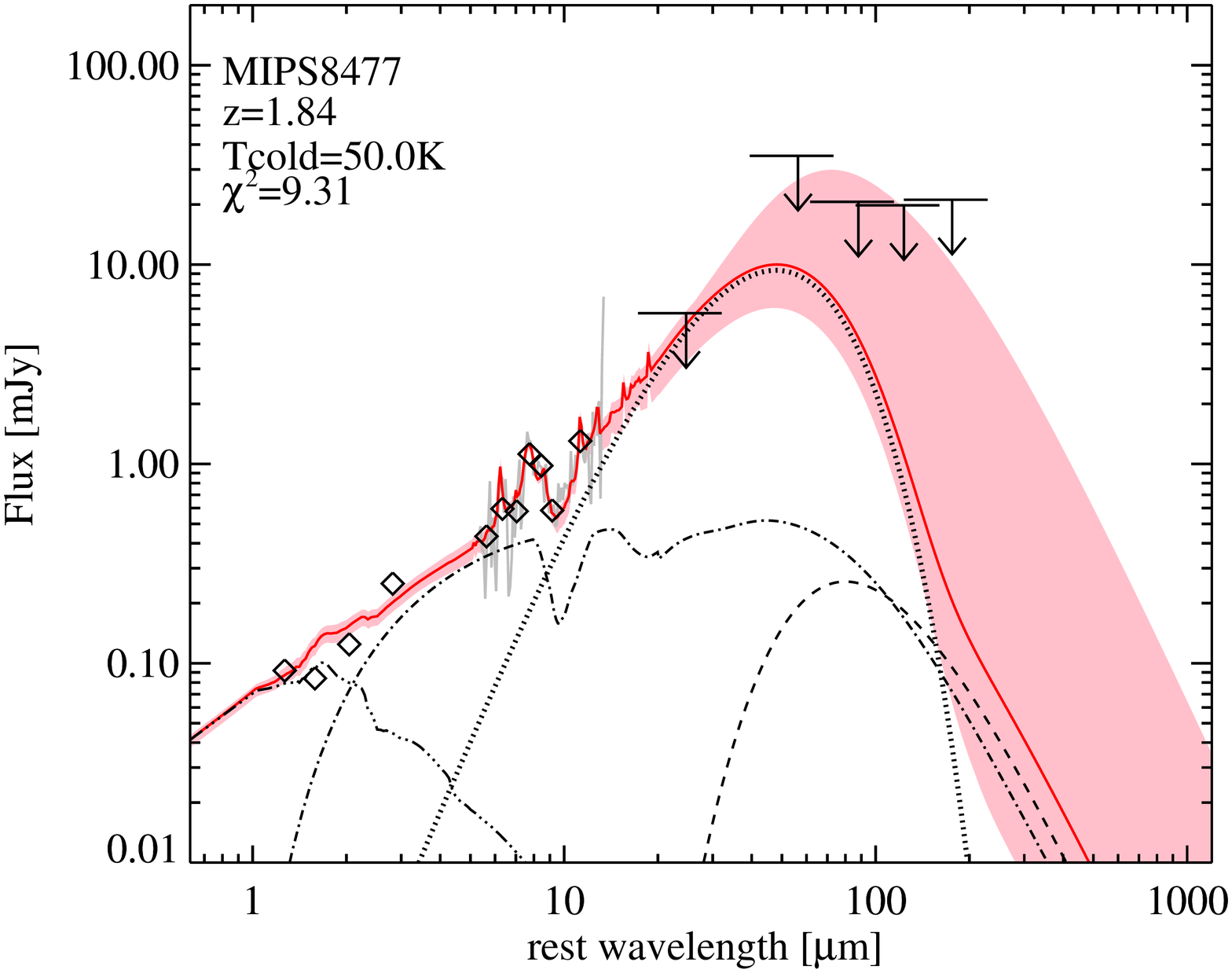}{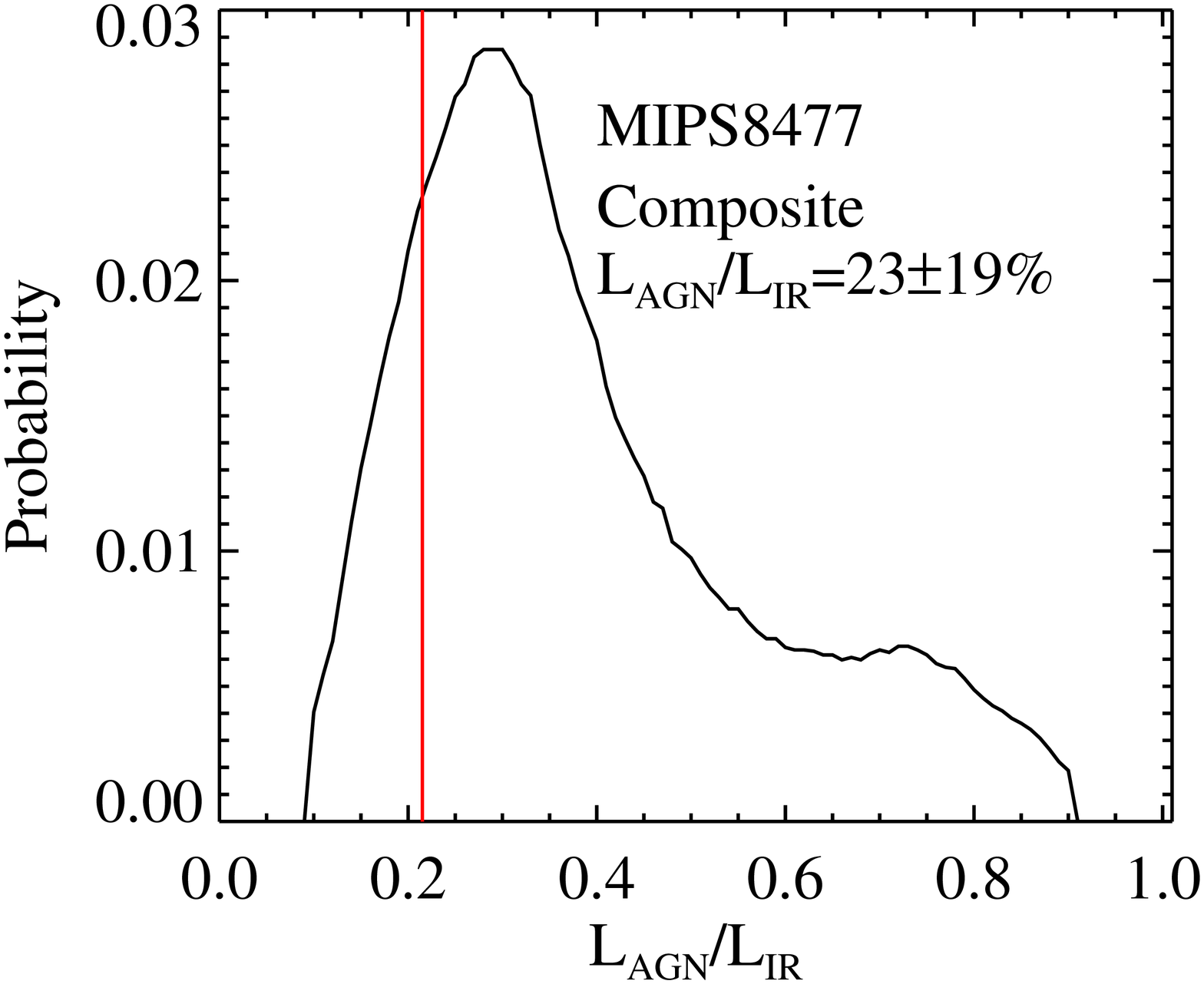}
\caption{Example of an MCMC-based $L_{\rm{AGN}}/L_{\rm{IR}}$ probability distribution for the composite source MIPS8477. To the left is shown the corresponding best-fit SED model.  This is an example where the uncertainty on the AGN fraction of $L_{\rm{IR}}$ is quite uncertain because of the lack of far-IR detections.  In the SED plot, the shaded region indicates the range of SEDs within $\chi^2$\,$<$\,$\chi^2_{min}+1$. The composite nature of this particular source is fairly secure however due to its having both strong PAH and a strong hot dust continuum. The vertical red line indicates the lowest $\chi^2$ solution. \label{fig_mcmc}}  
\end{figure}

We next consider the uncertainty on the AGN fraction of our sources, which directly translates into an uncertainty on their classification. Following the procedure described above, we construct posterior probability distributions for $L_{\rm{AGN}}$/$L_{\rm{IR}}$. From these posterior probability distributions, we compute the maximum likelihood $L_{\rm{AGN}}$/$L_{\rm{IR}}$ as well as its 68\%\ uncertainty.  Figure\,\ref{fig_mcmc} shows an example of this that shows a composite source, which has a strong uncertainty due to its lack of far-IR detections. This source also shows a hint of a secondary solution (at higher $L_{\rm{AGN}}/L_{\rm{IR}}$), which however has worse $\chi^2$ and moreover is disfavored for this source, where star-formation is already indicated by the strong PAH emission. Multiple peak solutions are found in $\sim$\,10\% of the sources\footnote{In these cases, we narrow the probability distributions by only taking $\chi^2$\,$<$\,$\chi^2_{min}+1$ solutions, which effectively isolates the best-fit peak}. From such probability distributions, we estimate the 68\%\ uncertainties on the AGN fraction, which vary from 1-50\%\, with a median of $\sim$\,7\% -- the higher uncertainty sources are typically associated with sources without far-IR detections, and where the far-IR upper limits are not very constraining (as the example shown). As another measure, ultimately of the classification uncertainty, we then compare the AGN fractions derived from the lowest $\chi^2$ solutions and the maximum likelihood ones derived from these MCMC-based probability distributions. The median difference between the two is 4\,$\pm$\,6\%. A comparison of classification based on either the lowest-$\chi^2$ or the maximum likelihood solution shows that the fraction of sources in each category is uncertain by $\sim$\,5\% (for the starbursts and AGN) and $\sim$\,10\% (for the composite sources). These are comparable to what we obtain by another method in Figure\,\ref{fig_templates}, suggesting that conservatively, we can state while individual source AGN fraction uncertainties can be much larger (though usually are not), the uncertainty on the fraction of sources within each category ("starburst", "composite" or "AGN") is in the range 5-10\%. These uncertainties on $L_{\rm{IR}}$ as computed above, as well as the maximum likelihood $L_{\rm{AGN}}/L_{\rm{IR}}$ values and their uncertainties are all given in Table\,\ref{table_lums}.

\begin{figure}[!h]
\plotone{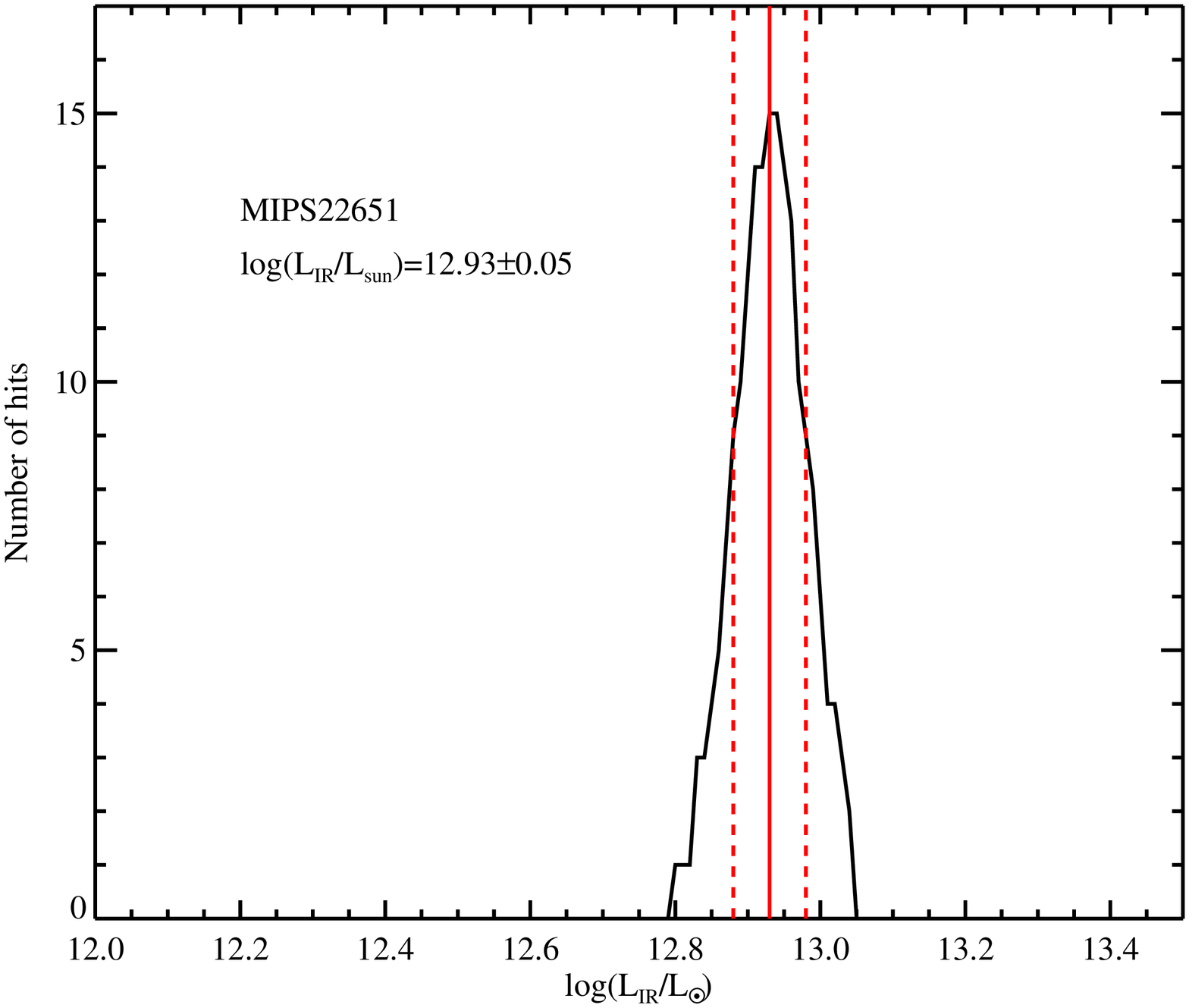}
\caption{Example of the posterior probability distribution on $L_{\rm{IR}}$ for the $z$\,$\sim$\,2 starburst, MIPS22651. The solid line shows the maximum likelihood value, with the dashed lines showing 1\,$\sigma$ errors on that (i.e. the range containing 68\%\ of the points).  \label{fig_lir_err}}
\end{figure}

\end{document}